# Characterization of Sb scaling and fluids in saline geothermal power plants: A case study for Germencik Region (Büyük Menderes Graben, Turkey)

Serhat Tonkul[a], Alper Baba[b,*], Mustafa M. Demir[c], Simona Regenspurg[d]

[*]Corresponding author: alperbaba@iyte.edu.tr

[a]Department of Energy Engineering, İzmir Institute of Technology, 35430, Gülbahçe, Urla, İzmir, Turkey

[b]Department of International Water Resources, İzmir Institute of Technology, 35430, Gülbahçe, Urla, İzmir, Turkey

[c]Department of Material Science and Engineering, İzmir Institute of Technology, 35430, Gülbahçe, Urla, İzmir, Turkey

[d]Helmholtz-Zentrum Potsdam, GFZ German Research Centre for Geosciences, Potsdam, Germany

## ABSTRACT

Turkey is located on the seismically active Alpine-Himalayan belt. Although tectonic activity causes seismicity in the Anatolian plate, it also constitutes an important geothermal energy resource. Today, geothermal energy production is heavily concentrated in Turkey's Western Anatolia region. Graben systems in this region are very suitable for geothermal resources. The Büyük Menderes Graben (BMG) is an area of complex geology with active tectonics and high geothermal potential power. Germencik (Aydın) is located in the BMG, where the geothermal waters include mainly Na-Cl-$HCO_3$ water types. This study examined the stibnite scaling formed in the preheater system of the Germencik Geothermal Field (GGF). The formation of the stibnite scaling on the preheater system dramatically reduces the energy harvesting of the GGF. Considering the stibnite scaling in the surface equipment, the optimum reinjection temperature was determined as 95 °C to prevent stibnite scaling in the GGF.

## 1. Introduction

Scaling in geothermal power plants is one of the important problems affecting the efficiency of the power plant. The most critical parameters that cause scale formation are water chemistry, temperature and, pH changes in the geothermal fluid during operation. Temperature and pH changes during operation affect mineral solubility and cause scale formation. While calcite and silica scales are common in most geothermal power plants, stibnite (Sb) scaling is rare. In power plants where calcite and silica scale types are observed, geothermal fluids are generally enriched in aluminum silicate, calcite, calcium sulphate, magnesium silicate, iron-magnesium silicate, while in power plants where stibnite scaling is observed, geothermal fluids are generally enriched in sulfide and sulfosalt minerals (Gunnlaugsson and Einarsson, 1989; Honegger et al., 1989; Kristmannsdóttir, 1989; Ölçenoğlu, 1986; Pátzay et al., 2003). Stibnite is an antimony mineral. Antimony occurs in trisulfide, $Sb^{3+}$, and pentasulfide, $Sb^{5+}$ in natural geothermal systems (Stauffer and Thompson, 1984; White, 1967). In some cases, antimony occurs as elemental Sb (Williams-Jones and Norman, 1997). Sb scaling has been addressed by many researchers worldwide. In New Zealand, Wilson et al. (2007) noted that Sb scaling developed in surface installations such as the heat exchanger at the Rotokawa geothermal station. Apart from New Zealand, this problem has been reported in an exploration well in Italy and in pipe equipment from the Berlin field, El Salvador (Cappetti et al., 1995; Raymond et al., 2005). Also, in Turkey, Sb scaling is the main handicap in many geothermal plants, including the western part of the Büyük Menderes Graben (BMG) system (Haklıdır and Balaban, 2019).

The BMG and the Gediz Graben are two important graben systems in Western Anatolia. The temperature of the geothermal fluid within these graben systems varies from region to region (Haklıdır and Şengün, 2020). The BMG is one of the most important graben systems which expands in the N-S direction and has a seismically active crust and is bounded by active E-W trending normal faults with a length of 150 km and a width of 10-20 km (Dewey and Şengör, 1979; Paton, 1992) (**Fig. 1**). On the other hand, the Gediz Graben is bounded by E-W and WNW-ESE trending normal faults with a length of 140-150 km and a width of 10-40 km (Bozkurt and Sözbilir, 2004; Seyitoğlu et al., 2000; Yılmaz et al., 2000). **Figure 2** shows a N-S cross-section of the BMG. As can be seen in **Fig. 2**, the graben has an asymmetrical structure. The Quaternary alluvium is bounded by normal faults with Pliocene and Miocene sediments. Miocene sediments were separated from the Menderes Metamorphics by detachment faults in the north and south of the BMG. This huge structure has several geothermal resources from the west to the east and can be considered a geothermal basin with a continued identification of new geothermal systems (Şimşek, 2015) (**Fig. 3**). There are more than 65 geothermal power units in Turkey, and most of them use the ORC-binary cycle. Sb-rich deposits have recently been observed in the GGF, particularly in Beştepeler, Pamukören and Maren geothermal power plants in Turkey (Kaypakoglu et al., 2015).

The Germencik (Aydın) region is well known for its high-temperature geothermal resources (Mutlu and Güleç, 1998). As a result of Curie point temperature calculations, Aydın et al. (2005) stated the depth of the heat source is 7-9 km in Western Anatolia, 30 km in Eastern Anatolia, and approximately 25 km in Central Anatolia. These results agree with reservoir temperatures in Western Anatolian graben systems (Koçak, 2012).

The purposes of this study are to determine the appropriate reinjection temperature for stibnite scaling observed in the surface equipment system and evaluate the hydrogeological and geochemical properties of the geothermal waters from the GGF. In this study, the stibnite scaling in the surface equipment system in the GGF was investigated in full detail. The types of scale that may occur in the geothermal wells are revealed, determining the hydrochemical properties of the geothermal waters. The reinjection temperature for stibnite scaling, which affects the power plant efficiency, was determined, and the behavior of Sb in the geothermal fluid was also investigated. These assessments were made using Phreeqc (Parkhurst and Appelo, 1999), WATCH (Bjarnason, 1994) speciation code, and Geochemist's Workbench software (Bethke, 2002).

## 2. Geological and hydrogeological settings in the study area

Germencik is an active geothermal area hosted by the Menderes Massif. Geothermal brine is found in a deep metamorphic reservoir at a depth of between around 2000 m and 3000 m, and a deep marble reservoir has been detected at a depth of 1440 m. Geothermal studies on the GGF have been ongoing since 1965. Because of the high geothermal potential of the GGF, the general geological, geophysical, hydrogeological, and environmental characteristics have been studied by Avşar and Altuntaş (2017), Correia et al. (1990), Doglioni et al. (2002), Filiz (1982), Gemici and Tarcan (2002), İlkışık (1995), Koçyiğit (2015), Khayat (1988), Seyitoğlu and Işık (2015), Şengör and Yılmaz (1981), Şimşek (1985). There are nearly 100 production wells in the Germencik region, and the reservoir of the geothermal fluid is located in different lithological units of the Menderes Metamorphic units (Özgür, 2018). The first geothermal well in the region was drilled by MTA (General Directory of Mineral Research and Exploration) in 1967, and exploration continued with deeper wells between 1982 and 1986. The reservoir temperature calculations made by MTA using different geothermometer applications showed that the reservoir temperature is between 190 and 232°C, and that the region has a high geothermal energy potential. These results are also supported by isotope results (Şimşek, 2003).

The study area has two main geothermal reservoirs. The first reservoir is located within the Neogene age sediments, which include conglomerates and sandstones. The second reservoir is located within the Menderes Massif of Paleozoic age. The Menderes Massif contains quartz schists, mica schists, gneiss, and marble. The thickness of the quartz schist and mica schist units varies between 600-800 m, while the thickness of the marble units is between 200-250 m. These two reservoirs are not directly connected and have different reservoir temperatures. The caprock formations for the geothermal system include the Neogene and Quaternary impermeable claystone and mudstone that fill the basin in this region. The main fault systems in the study area are located in the north and west of the site. Apart from the fault systems, basement units and cover rocks are spread geologically around the grabens. A detailed geology map of the study area is given in **Fig. 4.**

There are 25 geothermal wells in the study area, including 11 production wells, 9 reinjection wells, and 5 observation wells, and the depths of the wells vary between 1400-3500 m (**Fig. 5**). The 3D conceptual model of the study area was created in the Leapfrog Geothermal software, and well logs were used to create the model (**Fig. 5**). The depth of the shallowest well in the study area is 1449 m, and all these wells tap the thermal fluids from the Menderes Massif reservoir.

According to the 3D conceptual model, the geothermal waters in the study area are recharged from the high-altitude regions. The Gümüşdağı horst constitutes the main recharge area of the geothermal waters (**Fig. 5**). Geothermal waters move up to the surface thanks to the tectonic lines after heating at depth. In this way, a favorable environment is created for the geothermal systems in the region.

## 3. A brief definition of the Sb scale precipitation mechanism

Scale precipitation which causes a decrease in net energy production is the main issue in water-dominated geothermal systems (Haizlip et al., 2012). The hot geothermal fluid which interacts with the rocks at depth gains its chemical composition from the rocks. Therefore, under high temperature and pressure conditions, the geothermal fluid is enriched in minerals due to rock-water interaction. Hence, it is important to understand the relationship between reservoir rock and thermal water and interpret the scale forming mechanism that may persist.

Natural geothermal systems can have low concentrations of stibnite (Karpov, 1988; Krupp and Seward, 1990; Stauffer and Thompson, 1984; White, 1967; Zotov et al., 2003). As with other minerals, stibnite precipitation in geothermal systems depends on temperature and pH conditions. However, the precipitation of stibnite is almost impossible at high temperatures (above 350°C) due to its increased solubility in water (Zotov et al., 2003). There are two types of Sb species that occur in geothermal systems: hydroxide and sulfide complexes (Sorokin et al., 1988; Spycher and Reed, 1989). Sulfide complexes of stibnite have been evaluated at low temperatures by Babko and Lisetskaya (1956), Gushchina et al. (2000), Krupp (1988), and Mosselmans et al. (2000). Hydroxide and sulfide complexes in geothermal fluids are associated with the concentration of $H_2S$ in the brine. Krupp (1988) studied the formation of sulfide and hydroxide complexes in geothermal fluids containing $H_2S$ over a wide temperature range and suggested that in geothermal systems containing sulfide, sulfide components are dominant at temperatures below 100°C, while hydroxide components are dominant at higher temperatures.

Stibnite dissolves in water to form the "hydroxide", $Sb(OH)_3$ according to the following equation;

$$Sb_2S_3 + 6H_2O = 2Sb(OH)_3 + 3H_2S \qquad (1)$$

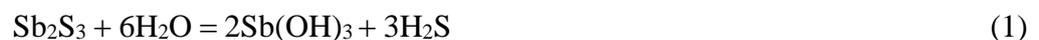

However, stibnite can also dissolve in geothermal fluids containing $H_2S$ and form sulfosalts which are known as sulfide complexes;

$$Sb_2S_3 + H_2S = H_2Sb_2S_4 \qquad (2)$$

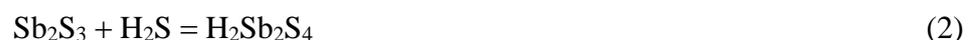

Sulfosalt forms of stibnite are essential when the temperature of the geothermal fluid is low in the presence of $H_2S$ (Brown, 2011).

Stibnite scaling depends on the type of power plant as well as on the reservoir rocks of the geothermal system. ORC-binary systems are suitable for power plants with low geothermal fluid temperatures (150-190°C) (DiPippo, 2012). The energy of the geothermal fluid in an ORC-binary system is transferred through the heat exchanger with the working fluid. The composition of the working fluid may be n-butane, n-pentane, or an ammonia-water mixture. The working fluid has a low boiling point, and it creates superheated steam conditions for the power plant (Stapleton et al., 2016). In ORC-binary systems, the steam in the turbine is used for electricity generation and returned to the condenser. The steam is converted into the working fluid in the condenser, and it returns to the heat exchanger at a lower temperature. When the higher temperature geothermal fluid encounters the low temperature working fluid in the heat exchanger, it transfers its heat to the working fluid, and the temperature of the geothermal fluid drops rapidly (**Fig. 6**). Also, the increase in the $H_2S$ concentration of the geothermal fluid causes the pH to decrease. Hence, surface equipment such as heat exchangers with low-temperature and low pH conditions (due to high $H_2S$ concentrations in the geothermal fluid) is equipment in which the suitable conditions for stibnite precipitation are created. Therefore, in addition to the calcite scaling in production wells, heat exchangers are also crucial for stibnite scaling in ORC-binary systems. Stibnite precipitation can also be observed in vaporizers and preheaters (Brown, 2011).

## 4. Methods
### 4.1. Geothermal water sampling

Geothermal water sampling was carried out in the GGF. Geothermal water samples were collected from the wells and the surface equipment system. A total of 11 geothermal water samplings were carried out, including from 9 production and 2 reinjection wells (**Fig. 6**). The samples were collected from the wellheads (before the separator) (**Fig. 6**). And, the geothermal waters were also collected from the preheater in and the preheater out (**Fig. 7**). 100 mL plastic bottles for heavy metal analysis, 250 mL plastic bottles for major-minor anions/cations and silica analysis, and 1 L plastic bottles for isotopes were used. After collecting samples for heavy metal analysis, they were acidified with 2% $HNO_3$. The aim is to prevent heavy metals from settling in the bottles. After collecting all the samples, the caps of the bottles were tightly closed, and contact of the samples with air was prevented. Physical and chemical parameters of the geothermal waters were analyzed for in the laboratories of the Izmir Institute of Technology. Major and minor component analysis was performed using ICP-MS (Inductively Coupled

Plasma Mass Spectrometry), and heavy metal analysis was performed using ICP-OES (Inductively Coupled Plasma Optical Emission Spectroscopy). The $SiO_2$ values for the geothermal waters were determined by the UV-spectrophotometric method with Hach-Lange DR5000. Isotope analysis of the geothermal waters was carried out in the State Hydraulic Works (DSİ) laboratories.

### 4.2. Rock and scale samples

Rock samples at different depths were collected from 4 geothermal wells, and the samples were analyzed in the laboratories of the Izmir Institute of Technology. XRD and XRF analyses of rock samples belonging to geothermal wells were evaluated. The aim here is to reveal the paths followed by geothermal waters while reaching the surface from the depths and the rock-water relationship. In addition, understanding the relationship between the sulfide-type scale in the GGF and reservoir rocks by XRD and XRF analysis is another goal.

In addition to the rock samples, scale samples were collected from the preheater system at the GGF, and XRD, XRF, and SEM analysis was performed.

X-ray crystallography (XRD), X-ray fluorescence spectroscopy (XRF), and scanning electron microscope (SEM) analyses of rock samples were evaluated in the laboratory of the İzmir Institute of Technology. XRF analysis of 4 rock core samples collected from geothermal wells (W_TR_002, W_TR_003, W_TR_007, and W_TR_009) was performed on the Spectro IQ II device. Spectro IQ II can deliver element concentrations from Sodium (Na-11) to Uranium (U-92) with high sensitivity at the ppm-level. Also, by using Philips X'Pert Pro, X-ray diffraction (XRD) analyses of rock samples were evaluated. XRD gives information about the concentration of phases, the amount of non-crystalline phases, and the crystal size of the material. By using FEI QUANTA 250 FEG scanning electron microscope (SEM), rock structures were visualized at micro size.

Information on rock samples collected from the geothermal wells at different depths is presented in **Table 1**.

### 4.3. Saturation indices for the geothermal waters

Silica, calcite, and sulfide scaling are the most important problems of geothermal power plants, reducing power plant efficiency. The water-mineral equilibrium state of geothermal fluids changes with temperature and partial pressure for each mineral from the depth to the surface.

The mineral equilibrium approach is based on calculating the saturation index for each temperature value and various minerals using the results of the chemical analysis of the fluid. To interpret this chemical concentration gained by the geothermal fluids, the saturation states of geothermal fluids according to various minerals were evaluated using the Phreeqc (Parkhurst and Appelo, 1999), and WATCH (Bjarnason, 1994) software. Mineral equilibrium states for the geothermal fluid at 10 different temperature values were investigated using the Phreeqc (Parkhurst and Appelo, 1999), and WATCH (Bjarnason, 1994) software. Also, mineral speciation evaluation of stibnite minerals at different temperatures and pH values was performed using Geochemist's Workbench software. Temperature and pH, major anions, and cations used as input parameters for the software are given in **Tables 2** and **3**, respectively. In Phreeqc, LLNL (The Lawrence Livermore National Laboratory) database was used, relying on the accuracy to be within ±0.5.

**4.4. Geothermometer applications**

The general purpose of geothermometers is to predict fluid temperature. As it is known, there are many different areas of use for hot water depending on the temperature at the surface. The hot water in the deep reservoir cools considerably until it reaches the surface and mixes with cold groundwater in different proportions. It is clear that the temperature of the deep fluid will be much higher than the temperature at the surface. If the path of the geothermal waters is short and the flow rate is high, their temperature is close to the reservoir rock temperature. Therefore, estimating reservoir temperatures with geothermometer methods is an important part of geothermal studies.

Some researchers have developed experimental geothermometers (Arnorsson, 2000; Arnorsson et al., 1983; Fournier, 1977; Fournier and Potter, 1982; Fournier and Truesdell, 1973; Fournier, 1992; Giggenbach, 1988; Kharaka et al., 1982; Kharaka and Mariner, 1989; Nieva and Nieva, 1987; Tonani, 1980; Truesdell, 1976).

The geothermometers used in the reservoir temperature calculations in the GGF are briefly explained below.

**4.4.1. Silica geothermometers**

Silica geothermometers are widely used in determining reservoir temperature, and they are based on the temperature-dependent solubility of silica in water. These geothermometers give good results at temperatures between 150-225°C (Fournier, 1977). Rapid silica precipitation is observed in the hot fluid moving from the aquifer towards the surface at higher temperatures. Therefore, if the reservoir whose temperature is above 225°C does not reflect the real temperature in the geothermal water.

**4.4.2. Cation geothermometers**

Cation geothermometers are geothermometers based on ion exchange. Ion exchange is a function of the reaction equilibrium constant (K), which depends on temperature. The ratio of ion-exchanged cation concentrations depends on the change of the equilibrium constant with temperature. Cation geothermometers give different reservoir temperatures, especially for hot water sources, due to processes such as mixing with cold-water and water-rock interaction affecting the chemical composition of the fluid during the rise of hot water.

Some silica and cation geothermometer relations proposed by various researchers are given in **Table 4**.

**5. Results**

**5.1. Hydrogeochemical properties of the geothermal waters**

**5.1.1. Physical properties of the geothermal waters**

The pH values of the geothermal waters in the GGF are between 6.7 and 8.54. It is thought that the pH values of the geothermal waters in the GGF between 6.7 and 8.54 may be due to the effect of dissolved $CO_2$. The source of $CO_2$ in the geothermal waters is possibly the thermal metamorphism of limestone in the reservoir. Electrical conductivity values vary between 5697 μS/cm and 6147 μS/cm. Total dissolved solids (TDS) in the geothermal fluids range from 3700 to 4200 ppm. The outflow temperatures of production wells are between 125°C and 148°C, and about 64°C in the reinjection wells. The highest outflow temperature is in well W_TR_009. The physical properties of the geothermal waters are given in **Table 2**.

**5.1.2. Chemical properties of the geothermal waters**

The chemical properties of geothermal waters are given in **Table 3**. Based on chemical analyses, the geothermal waters in the GGF are of the Na-Cl-$HCO_3$ water type. Na-Cl-$HCO_3$ waters are located in marble, quartzite, and mica-schist units, which are reservoir rocks for the geothermal waters.

As can be seen in **Table 3**, Sodium ($Na^+$) and Chloride ($Cl^-$) concentrations are relatively higher than those of other chemical constituents. $Na^+$ is the major cation in the geothermal waters of the GGF, with a concentration range from 1072 to 1355 mg/l.

The $Na^+/K^+$ ratio in the geothermal waters is a good indicator of the paths followed by the waters. A low $Na^+/K^+$ (<15) ratio means a high temperature in the geothermal waters. A low $Na^+/K^+$ ratio means that geothermal waters rise rapidly from the depths to the surface and do not have enough time to cool. The ratio of $Na^+/K^+$ in the GGF is 20 and high. Therefore, the geothermal waters rise to the surface slowly, transfer heat to the surrounding rocks, and cool down. This situation indicates the presence of near-surface reactions and conductive cooling in the geothermal waters. As salinity increases in the geothermal waters, the $Ca^{2+}$ values also increase. The $Ca^{2+}$ and $HCO_3^-$ values in the geothermal waters are associated with marbles in the Menderes metamorphic rocks. Silica ($SiO_2$) values are possibly due to the quartz mineral found in rocks such as quartzite and schist. Anion and cation distributions in the GGF are given in **Fig. 8**. As can be seen from the Pie diagrams, $Na^+$ is the dominant cation, $HCO_3^-$ and $Cl^-$ are the dominant anions.

Piper diagrams are triangular diagrams used to classify waters and express water facies (Piper, 1944). It is seen that the geothermal waters show similar facies characteristics in the Piper diagrams. According to the Piper diagrams, the GGF water samples are in the bicarbonate water group, and they are considered peripheral waters (**Fig. 9a** and **Fig. 9b**).

The Schoeller diagram shows the major and minor ions of water samples on a single graph (Schoeller, 1967). The semi-logarithmic Schoeller diagram (**Fig. 10a** and **Fig. 10b**) shows that the geothermal water samples have a similar composition in the GGF. It is seen that the concentrations of $Mg^{2+}$ and $SO_4^{2-}$ in the geothermal waters decrease, and they are enriched in $Ca^{2+}$, $Na^+$, $K^+$, $Cl^-$, and $HCO_3^-$ (**Fig. 10a** and **Fig. 10b**).

From the Piper and Schoeller diagrams for the water samples collected from the geothermal wells and surface equipment system, it is seen that there is no significant change in the water composition of the geothermal water from the production to the preheater system.

In geothermal waters, $Cl^-$ and B elements are essential in understanding the water-rock interaction (Aggarwal et al., 2000). The chloride ($Cl^-$) concentrations ranging from 1183 to

1357 mg/l in the GGF, which is located 20 km away from the Aegean Sea. Karakuş and Şimşek (2013), and Güner and Yıldırım (2005) explained that the Cl$^-$ concentrations are associated with a marine sediment effect during the Pleistocene in the GGF. The same authors stated that the effect of seawater intrusion on GGF is 7-8%. Aggarwal et al. (2000) stated that high Cl/B ratios might indicate a seawater effect, while low Cl/B ratios are associated with magmatic volatiles. Baba and Sözbilir (2012) stated that geothermal waters could gain their chloride concentrations from hot water-related faults. Filiz et al. (2000) explained that boron values in geothermal waters are associated with mantle or marine sediments undergoing metamorphism.

Germencik geothermal waters are rich in chloride (Cl$^-$) and modest in boron (B) concentrations. Boron concentrations range from 36 to 50 mg/l in the GGF. The Cl/B ratio for the GGF is given as 28.10, and there is a relationship between Cl and B concentrations. This linear relationship between chloride and boron concentrations indicates that the geothermal waters are recharged from the same deep reservoir (**Fig. 11**). The boron in the geothermal waters is probably not derived from the sedimentary rocks and possibly from metamorphic ones such as gneiss and schist (**Fig. 11**).

### 5.1.3. Heavy metals and trace elements in the geothermal waters

Heavy metal and trace element analyses of geothermal waters in the GGF were evaluated. As can be seen in **Fig. 12**, Aluminum (Al) and arsenic (As) values in the geothermal fluids are between 41-233 and 13-153 ppb, respectively. The lowest Al and As concentrations were observed in well W_TR_006. Boron (B) values in the geothermal well fluids are close to each other and average 46 ppm. The boron concentration in the geothermal waters is due to boron-containing minerals such as mica and feldspar of the Menderes Massif metamorphics containing pegmatitic tourmalines. Lithium (Li) concentrations in the geothermal waters are also close to each other, and an increase is seen in those of well W_TR_006. In the geothermal water of the GGF, Li$^+$ is associated with secondary processes. Therefore, it is used as a trace element. Lithium concentrations of the geothermal waters range from 4.96 to 7.04 ppm. Sb concentrations in the geothermal waters range from 0.94 to 1238 ppb. As seen in **Fig. 12**, the highest Sb concentrations were observed in wells W_TR_003 (1238 ppb), W_TR_005 (692 ppb), and W_TR_008 (612 ppb). In the surface equipment system, Sb concentrations increase at the entrance of the preheater.

### 5.1.4. Stable isotopes

Oxygen and hydrogen isotope studies were carried out to understand the evolution of geothermal waters in the GGF. O is the most abundant element in the earth's crust and is generally found in higher amounts in rock reservoirs (Clark and Fritz, 1997). Unlike O, H is generally found in waters rather than minerals and rocks. The different behaviour of these two elements is important for the isotopic evaluation of waters in high-temperature systems (Clark and Fritz, 1997).

The $\delta^{18}O$ isotope values vary between −1.69‰ and −2.05‰, and the $\delta D$ isotope values vary between −38.07‰ and −40‰ in the thermal waters (**Fig. 13a**). The stable isotope results show that geothermal waters have a meteoric origin in the GGF and suggest a substantial $\delta^{18}O$ isotope enrichment in the thermal waters in the Germencik region. This means that geothermal waters have a longer contact time with reservoir rocks at depths. This is also consistent with the slow rise of the geothermal waters to the surface. As a result of the interaction between the rocks with heavy isotopes and the geothermal waters, the geothermal waters are enriched in heavy isotopes, and their composition changed towards positive $\delta^{18}O$ values (**Fig. 13a**). Therefore, the high oxygen concentration may indicate that the geothermal reservoir temperature is high. The geothermal waters show a strong oxygen isotope shift in **Fig. 13a,** indicating exchange reactions between water and rock at over 220°C. In other words, this supports that high-temperature geothermal systems exit in this region and the high dissolution of Sb in the geothermal system. Another reason for geothermal waters enriched in $\delta^{18}O$ may be the carbonate units containing marble in the Menderes Metamorphic rocks. Since $\delta D$ is mostly found in oceans and natural waters, there are more negative values in the geothermal waters (**Fig. 13a**).

The most widely used radioactive isotope in hydrogeochemical studies is tritium ($^3H$) (Dansgaard, 1964). Due to its radioactive nature, it is used for determining the age of groundwater. In addition, the duration of the groundwater in the reservoir is estimated. The low $^3H$ (<5) content of the geothermal waters in the basement rocks indicates that the geothermal waters are older than 60 years (**Fig. 13b**). Also, the relationship diagram between tritium ($^3H$) and electric conductivity shows that geothermal fluid has a deep-water circulation (**Fig. 13b**).

## 5.2. Evaluation of Rock and Scale Sample

### 5.2.1. Rock samples

Rock samples from different depths were collected from 4 geothermal wells in the GGF. Here, the aim is to reveal which rocks in the reservoir may be associated with the stibnite scaling observed in the preheater system. The mineralogy and elemental compositions of the rock samples were evaluated, and the mineral compositions of the reservoir rocks were revealed. These 4 wells from which rock samples were collected; W_TR_002, W_TR_003, W_TR_007, and W_TR_009. W_TR_002 and W_TR_003 are located in the south of the study area, whereas W_TR_007 and W_TR_009 are located in the north of the study area (**Fig. 14**).

In well W_TR_002, alluvium units are between 0 and 25 m. At 25-905 m, sedimentary units are observed, while Menderes Metamorphic units are observed below 905 m. Rock samples were collected between 2325 and 2335 meters from well W_TR_002. As seen in the W_TR_002 well log, there are mica schist units at these depths. As a result of XRD analysis of samples from well W_TR_002 from a depth of between 2325 and 2335 m, feldspars (albite ($NaAlSi_3O_8$)) and (anorthite ($CaAl_2Si_2O_8$)), silica (quartz ($SiO_2$)), and iron minerals (pyrite ($FeS_2$) and magnetite ($Fe_3O_4$)) were detected. In addition, another mineral found in schists, antimony (Sb), which is one of the sulphide minerals, was detected between these depths. (**Fig. 14**).

In well W_TR_003 with a depth of 3074 m, rock samples were collected from 2995 m. This depth corresponds to the transition zone between the mica-schist units and the marbles. As seen as a result of XRD analysis, carbonate minerals (calcite ($CaCO_3$) and dolomite ($CaMg(CO_3)_2$)) associated with marbles were found at this depth. Feldspar (albite ($NaAlSi_3O_8$)) and silica minerals (quartz ($SiO_2$)) contained in mica schists were also found at this depth. Diffraction signals at $2\theta = 15°$ and $32°$ were assigned to antimony (Sb) in well W_TR_003 (**Fig. 14**).

In well W_TR_007 with a depth of 2451 m, there is alluvium between 0-35 m and sedimentary units between 35-857 m. Metamorphic units are observed at 857 m below. In well W_TR_007, rock samples were collected between 1265-1285 m, and these depths correspond to calcschist units. In W_TR_007 well, muscovite, one of the mica group minerals, was detected, unlike in other wells. Enrichment of geothermal waters in muscovite minerals ($KAl_3Si_3O_{10}(OH)_2$)) is attributed to the dissolution of potassium feldspars by geothermal waters. Diffraction signals at

2$\theta$ = 28) indicate antimony (Sb) at this depth in W_TR_007. Other than that, their mineral composition is similar to that of wells W_TR_002 and W_TR_003.

Well W_TR_009 with a depth of 2568 m has alluvium between 0-35 meters. 35-500 m sedimentary units and below 500 m metamorphic units are observed in this well. Rock sampling was made between 900 and 920 m for this well. In the well log, mica schist units were observed between these depths. Minerals seen in the XRD pattern of well W_TR_009 have similar characteristics to those the other wells. Muscovite ($KAl_3Si_3O_{10}(OH)_2$)), a mica group mineral, was detected in diffraction signals (2$\theta$) at 9°, 32°, and 62°, respectively. Diffraction signals at 22° and 28° indicate feldspar minerals. Diffraction signals at 21°, 39°, 51°, and 68° indicate the presence of quartz minerals in the well W_TR_009.

The rock samples collected from different depths were subjected to elemental analysis using XRF. The chemical compositions of all the rock samples by XRF were evaluated with respect to different depths. **Fig. 15** shows that the elemental compositions of the collected rock samples are enriched in Si, Al, Ca, Fe, K, and Sb. In the wells in the rock samples collected from the north of the GGF, Si values increase, while Al values decrease. The increase/decrease of the element is related to the reservoir rocks at the depth of the rock sample. For example, Ca, Si, and Al values increased in rock samples taken from depths close to the schist marble contact. On the other hand, an increase was observed in Si, Al, Sb, and Fe values in rock samples collected from depths close to gneiss and schist units. The findings obtained from the XRD and XRF results show that the Sb scale observed in the preheater system in the GGF may be related to the gneiss and schist units of the Menderes Metamorphic Massif.

### 5.2.2. Stibnite scale sample in the preheater system

The Sb scale in the preheater system was subjected to elemental analysis using XRD, XRF and SEM. **Fig. 16a** shows the photographic image of the Sb deposit in the preheater. The red color of the deposit, which is composed of a fine powder, is evident. The photographic image indicates that the scale contains reddish-orange amorphous stibnite. The thickness of the Sb deposit is between 1 and 4 mm (**Fig. 16b**).

The diffractogram has a broad signal centered around 25° and a very broad reflection around 52°. The XRD pattern shows amorphous particles, which means there is no crystalline reflection for stibnite (**Fig. 17a**).

SEM analysis was also performed to see the morphology of the scale. It shows heterogenous spherical colloids whose size is 1.1 ± 0.5 μm (**Fig. 17b**). This type of morphology is quite similar to that of amorphous stibnite particles that have been reported previously many times in the literature (Çiftçi et al., 2020; Gill et al., 2013; Pan et al., 2009; Zakaznova-lakovleva et al., 2001).

The elemental composition of the scales shows that Sb and S are mainly observed at 56.57% and 21.87%, respectively (**Fig. 18**). These ratios are compatible with stibnite stoichiometry in terms of the Sb/S ratio, which is 2.5. The Sb/S ratio varies depending on the sulphide concentration, and it is consistent with the ratios in the literature (Spycher and Reed, 1989; Zotov et al., 2003).

## 5.3. Reservoir temperatures of the GGF
### 5.3.1. Silica geothermometer results

Within the scope of the study, $SiO_2$ concentrations (as ppm) were obtained from the chemical analysis results evaluated in the GGF. $SiO_2$ concentrations range from 125 to 165 ppm in the GGF. The reservoir rock temperatures calculated for the geothermal waters representing the geothermal aquifer with the silica geothermometer equations are presented in **Table 5**. As can be seen in **Table 5**, the reservoir temperatures calculated according to silica geothermometers vary between 139°C and 167°C.

### 5.3.2. Cation geothermometer results
#### 5.3.2.1. Na-K geothermometers

Na/K geothermometers are based on cation exchange reactions between albite ($NaAlSi_3O_8$) and K feldspars (Fournier and Potter, 1979). The cation exchange reaction is given Eq. 3.

$$NaAlSi_3O_8 + K^+ \Leftrightarrow KAlSi_3O_8 + Na^+ \text{ (K-feldispar)} \qquad (3)$$

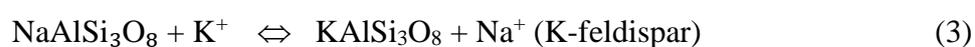

The geothermal waters with a high chloride concentration from high-temperature reservoirs (>180°C) are suitable for this type of geothermometer. However, it can also be applied at low temperatures where geothermal water remains longer in the reservoir. Since these geothermometers depend on the ratio of $Na^+$ and $K^+$, they are less affected by dilution and steam loss due to the cold-water mixture. Therefore, in systems with cold water mixtures, they give more accurate results than silica geothermometers (Fournier and Potter, 1979). The Na/K geothermometer results show that a reservoir temperature for the geothermal fluid in the GGF ranges from 146 to 188°C (**Table 5**).

### 5.3.2.2. Na-K-Mg geothermometers

The Na-K-Mg triangular diagram was proposed by Giggenbach (1988). This diagram consists of a combination of Na/K and K-Mg geothermometers. As can be seen from the Giggenbach diagram (**Fig. 19**), the geothermal waters in the GGF tend to approach the Na and K line. This is related to the Menderes Metamorphic Massif, including gneiss and schist units. These metamorphic rocks are enriched in feldspar minerals ($NaAlSi_3O_8$ and $KAlSi_3O_8$). According to the Giggenbach triangular diagram, it is seen that the geothermal waters in the GGF are in a partial equilibrium with reservoir rocks (**Fig. 19**). Therefore, Na/K geothermometers can be applied to the geothermal water samples of the GGF. The cation temperatures shown in **Table 5** agree with the temperatures estimated by the Giggenbach diagram (**Fig. 19**). Giggenbach's diagram suggests a deeper or longer circulation for the GGF.

The reservoir temperature modelling was also performed for the GGF using wells that have static temperature data. According to the numerical and block diagram obtained, the highest temperatures are in the north of the GGF (**Fig. 20a** and **Fig. 20b**). As seen in the numerical model, reservoir temperatures in the north of the geothermal field vary between 150°C and 200°C. These values agree with the reservoir temperatures obtained by geothermometer calculations.

The reservoir temperature of the geothermal fluids in the BMG varies from region to region. The reservoir temperature in the western end of the BMG can reach 276°C, while it can reach 245°C at the eastern end. In the middle part of the BMG, the reservoir temperature of the geothermal fluid is around 169-188°C (Haklıdır and Şengün, 2020). Hence, geothermometer calculations agree with the actual reservoir temperature in the GGF.

## 5.4. Mineral precipitations in the GGF

### 5.4.1. Scale in the geothermal wells

For the GGF, saturation indices were calculated at different temperatures from 20°C to 200°C using the chemical models Phreeqc and WATCH for the geothermal fluids, where reservoir temperatures are lower than 190°C. The geothermal fluids are supersaturated with quartz and chalcedony minerals in all wells below 140°C and 120°C, respectively (**Fig. 21**). Otherwise, the geothermal fluids are undersaturated with amorphous silica above 40°C. However, as the temperature decreases (<40°C), amorphous silica precipitation can be expected. The formation of silica forms in geothermal systems is strongly dependent on pH and temperature. Silica formation can also vary according to the amount of silica dissolved in the reservoir and the type of power plant (Brown, 2013; Demir et al., 2014; Utami, 2011, Zarrouk et al., 2014).

The geothermal fluid is under-saturated with respect to the mineral anhydrite ($CaSO_4$) at all temperatures. Therefore, there is no risk of anhydrite precipitation in the geothermal wells. The geothermal fluids are supersaturated with respect to calcite. However, W_TR_001, W_TR_002, W_TR_009, and W_TR_010 are under-saturated with respect to calcite below 100°C. As the amount of $CO_2$ dissolved in the geothermal fluid increases, the amount of dissolved calcite increases. Therefore, as the geothermal fluid rises to the surface, the decrease in the amount of $CO_2$ will cause calcite precipitation with the decrease in pressure (Brown, 2013; Tarcan et al., 2016).

### 5.4.2. Scales in the preheater system

For the preheater system in which stibnite scaling is observed, the saturation index model at the preheater in and the preheater out was implemented. The physical properties of the geothermal fluid vary from production to reinjection. In particular, the temperature of the geothermal fluid decreases from the production well until it reaches the power plant. The inlet temperature of the geothermal fluid in the preheater is 80°C, while the outlet temperature of the geothermal fluid in the preheater is 65°C. **Fig. 22** shows that the geothermal fluids are supersaturated with stibnite (Sb) minerals below 90°C in the preheater system. Wollastonite ($CaSiO_3$) and anhydrite ($CaSO_4$) are under-saturated at all temperature values. Amorphous silica ($SiO_2$) and fluorite ($CaF_2$) in the preheater system are under-saturated at all temperature values like wollastonite and anhydrite. Calcite ($CaCO_3$) solubility decreases with increasing temperature. Therefore, calcite scaling would not be expected in the low-temperature preheater system. Considering the inlet and outlet temperatures of the preheater system, silica and calcite-based scaling should not be expected in the GGF.

**5.5. Thermodynamic modelling for Sb in the geothermal waters**

Geothermal fluid changes thermodynamically from depth to surface conditions with changes in pH, temperature, and pressure. These thermodynamic changes depend on water-rock interaction and the non-condensable gases (NCG) that the geothermal fluid contains. Therefore, it was aimed to obtain thermodynamic information about Sb speciation under different temperature and pH conditions to observe the antimony behaviour in the hydrothermal system of the GGF.

Water-rock interactions in the thermal waters producing stibnite species were investigated using activity diagrams. In the activity diagram, according to the dissolved $H_2S$ and Sb concentration in the geothermal fluid, the Sb species that can occur at different temperatures and pH values in the geothermal fluid were revealed. In Germencik geothermal waters, the $H_2S$ level in the gas phase is 0.21% (Haizlip et al., 2013). Stibnite concentrations in the brine are given in **Fig. 12.**

Temperature values between 50 and 200°C were used to interpret the precipitation mechanism of the stibnite with pH and temperature because the solubility of the stibnite in the geothermal fluid is constant at higher temperatures (>200°C) (Zakaznova-Iakovleva et al., 2001). As the temperature and pH values decrease, the predominance of stibnite in the geothermal fluid increases. This means that stibnite precipitates at low temperature and pH values.

Based on the thermodynamic model, the solubility of the stibnite is controlled by the thioantimonite species between 50 and 90°C in the presence of $H_2S$ (**Fig. 23**). Thioantimonite species begin to dominate at 90°C. Acidification takes place in the geothermal fluid due to the oxidation of hydrogen sulfide ($H_2S$) in the pink rectangle region (**Fig. 23**). The red dashed ellipses were used to show the emergence of different stibnite species at low (50°C) and high temperatures (200°C). Acidification of the liquid causes a dramatic decrease in solubility and leads to stibnite precipitation. Therefore, in the dominant region of thioantimonite species (pink rectangle at 90°C), pH has a significant effect on stibnite solubility. Hydroxothioantimonite species become more dominant at higher temperatures in the brine (>150°C) (**Fig. 23**).

According to the model, in addition to the red amorphous stibnite (metastibnite) in the preheater system, it was concluded that different stibnite species might occur. Findings obtained from both the saturation index and the thermodynamic modelling confirm that stibnite precipitates in low-temperature surface equipment such as the preheater.

## 6. Discussion

The BMG, which can be considered as a geothermal basin, is tectonically one of the largest graben basins extending from east to west with a length of 150 km. The most important characteristic of the graben is that it has an asymmetrical structure. This huge structure has various hot water springs, hidden geothermal areas, and many geothermal power plants along a length of 150 km (Tezcan, 1979). Since there are many high-medium enthalpy geothermal fields in the graben, it provides suitable conditions for geothermal power plants. Geothermal power plants in Turkey are generally binary-cycle and flash-type power plants, and the power plant type varies according to the reservoir temperature. Binary cycle power plants are common worldwide as well as in Turkey (Bertani, 2016; DiPippo, 2012). The reason why binary cycle power plants are widely used is that they are easy to maintain and install operationally. However, the geothermal fluid undergoes physical and chemical changes thermodynamically during operation, causing scale problems that reduce the power plant efficiency. Depending on the temperature and pressure changes in the reservoir, the geothermal fluid whose physical and chemical properties change causes scaling problems in wells, pipelines, and surface equipment systems in binary-cycle power plants. Scale problems seen in binary cycle power plants in Western Anatolia have been addressed by various researchers (Baba et al., 2015; Çelik et al., 2017; Demir et al., 2014; Yıldırım and Yıldırım, 2015).

The most important scale types in geothermal power plants are calcite, silica, and sulphide scales. Calcite is the most abundant compound in these geothermal waters. Its solubility increases with the partial pressure of $CO_2$ and decreases with temperature. Therefore, the partial pressure of $CO_2$ and temperature are the most important parameters affecting calcite solubility. This thermodynamic feature of calcite causes efficiency loss by clogging the production wells and transport lines in geothermal power plants. Silica is found in geothermal waters in the form of amorphous silica, chalcedony, and quartz and the solubility of silica depends on temperature and pH. As the temperature decreases, the amount of dissolved silica decreases. Therefore, silica scaling takes place in reinjection wells and heat exchangers in geothermal power plants. Sulphide species can cause scale problems in all geothermal fields with low, medium, and high enthalpy. Among the sulphide species, stibnite scaling is a common problem in geothermal fields. Stibnite solubility is controlled by pH, temperature, and $H_2S$ concentration. Increasing $H_2S$ concentration in geothermal fluid causes pH decrease, and stibnite scaling starts with decreasing temperature. Therefore, stibnite scaling is seen in low-temperature surface equipment such as heat exchangers, preheaters, and condensers in geothermal power plants.

The Germencik geothermal power plant is a binary cycle plant located in the west of the BMG. The reservoir rocks of the GGF are composed of various Paleozoic aged schists (quartzite, quartzite-schists, calcschists, graphite-schists, and marbles) and gneisses. The GGF has two reservoirs: (1) the first is a Neogene aged conglomerate and sandstone reservoir that is fractured by faults, (2) the second and deep reservoir's rocks of the GGF are composed of Paleozoic aged various schists (quartzite, quartzite-schists, calcschists, graphite-schists, and marbles) and gneisses.

According to the Cl/B relationship obtained, the geothermal waters are recharged from the same deep reservoir and show the Na-CI-$HCO_3$ water type. Geothermal waters presenting the Na-CI-$CO_3$ water type come from gneiss, schists, and marbles in the Menderes Metamorphic Massif.

The main problem in the GGF is the stibnite scaling in the preheater system. In this study, the scale problems in the GGF have been investigated, dividing them into two groups: (1) Possible types of scale that may occur in the geothermal wells, (2) stibnite scaling in the preheater system

## 6.1. Possible mineral precipitation tendencies downhole

It is interpreted that there is different mineral precipitation from production to reinjection at different temperatures in the GGF (**Fig. 24**). According to **Fig. 24**, calcite precipitation can be expected in the production wells. While the geothermal fluid flows from the reservoir through the entire system in the GGF, there is no risk in terms of amorphous silica precipitation between 95°C and 195°C. However, as the temperature decreases, amorphous silica precipitation can be expected. Therewith, chalcedony, and quartz may precipitate below 135°C. No stibnite was detected in the production wells of the GGF.

## 6.2. Stibnite scaling in the preheater system

Analysis of rock samples collected from geothermal wells at different depths in the GGF showed that the reservoir rocks contain a Sb mineral. Especially, a Sb mineral was detected in XRD analysis of rock samples collected from depths close to gneiss and schist contacts. In addition, the highest concentrations in terms of elemental composition in rock samples belong to Sb. According to the rock sample analysis, it is understood that the origin of the stibnite scaling observed in the preheater system is gneiss and schists in the reservoir rocks. These results are consistent with the studies of other researchers. Dost (2018) explained that the stibnite source in the Germencik geothermal fields might be associated with metamorphic rocks such as gneiss and mica-schist. In a similar study in South Africa, Pearton and Viljoen (1986) stated that the source of stibnite is quartzite and quartz-muscovite schist.

In the XRD and SEM analysis performed on the scale sample collected from the preheater system, stibnite was found to be in the form of heterogeneous spherical colloids with amorphous particles. The reddish color in the photographic picture of stibnite and its amorphous morphology indicates that it is metastibnite. The low Sb/S ratio (2.5) in the elemental composition of the scale sample is consistent with many studies in the literature in terms of stibnite stoichiometry (Spycher and Reed, 1989; Zotov et al., 2003).

From the speciation thermodynamic diagram of stibnite, it is seen that different stibnite forms can be formed together with metastibnite in the preheater system. According to the thermodynamic model, it is predicted that thioantimonite ($H_2Sb_2S_4$) species may be present in the system with a decrease in temperature and pH. Brown (2013) attributed the formation of thioantimonite stibnite species to the presence of $H_2S$ in geothermal systems. The same author stated that increasing $H_2S$ concentrations would lead to a decrease in pH, and with the rapid decrease in temperature, stibnite scaling will take place. The $H_2S$ concentrations in the GGF fluids are high compared to those in other geothermal fields in the BMG and are 0.21% (Haizlip et al., 2013). Also, Osborn et al. (2007) attributed the low pH of the geothermal waters in the GGF compared to that in other geothermal fields in the BMG to marbles.

In the GGF, stibnite formation starts at 90°C in the preheater system. The geothermal fluid enters the preheater system at 80°C, and at this temperature, the geothermal fluid is supersaturated with respect to stibnite. (**Fig. 24**). **Fig. 24** shows that the calculated reinjection temperature is around 95°C to prevent stibnite scaling in the preheater system for the GGF. The results obtained show that stibnite precipitates in low temperature and pH conditions, and these results are consistent with those of many studies in the literature. Wilson et al. (2007) stated that stibnite precipitation is a major problem at the Rotokawa and Ngawha power stations. It has been stated that pH and temperature changes in the Rotokawa and Ngawha plant fluids is the main reason for stibnite precipitation in the preheater system. According to Wilson et al. (2007), a temperature lower than 100°C and pH < 8 provide optimum conditions for stibnite precipitation. In another study, Brown (2009) stated that temperatures lower than 90°C and pH < 9.7 are the best conditions for stibnite precipitation.

## 7. Conclusions and final remarks

In geothermal power plants, different types of scaling may take place in geothermal wells and surface equipment due to temperature, pH, and pressure changes in the geothermal fluid. The types of scaling vary according to the reservoir rocks and power plant type. Especially in the binary cycle power plants, the geothermal fluid entering the heat exchanger or preheater system directly after the production well causes the temperature of the fluid to drop rapidly. Heat exchanger and preheater systems of the binary cycle plant act as a sink, creating optimum conditions for stibnite scaling. Stibnite scaling causes efficiency and economic losses in the power plant by clogging the heat exchanger tubes. Therefore, determining the correct reinjection temperature is important in terms of potential scale formation.

Periodic mechanical cleaning is performed with high-pressure water jets, which is the most effective method to control stibnite scaling in the GGF. However, since this method requires the power plant to be completely shut down, it is very time-consuming and, at the same time, means loss of production. Although research and chemical companies are currently working on different chemical development studies for the prevention of stibnite scaling, the most suitable method is to use inhibitors before preheating by dosing appropriately. Caustic (NaOH) dosing can be an alternative for this. Therefore, caustic soda can be used at a certain dosing depth in production wells to prevent Sb scaling in the power plant.

**Acknowledgments**

The authors are grateful for the funding received from the European Union's Horizon 2020 research and innovation programme under grant agreement No. 850626 (REFLECT). The authors are also grateful to the two anonymous reviewers for their valuable contribution to improving our manuscript. The authors also gratefully thanked Dr. Christopher Bromley and Dr. Halldór Ármannsson for the detailed edition of the manuscript.

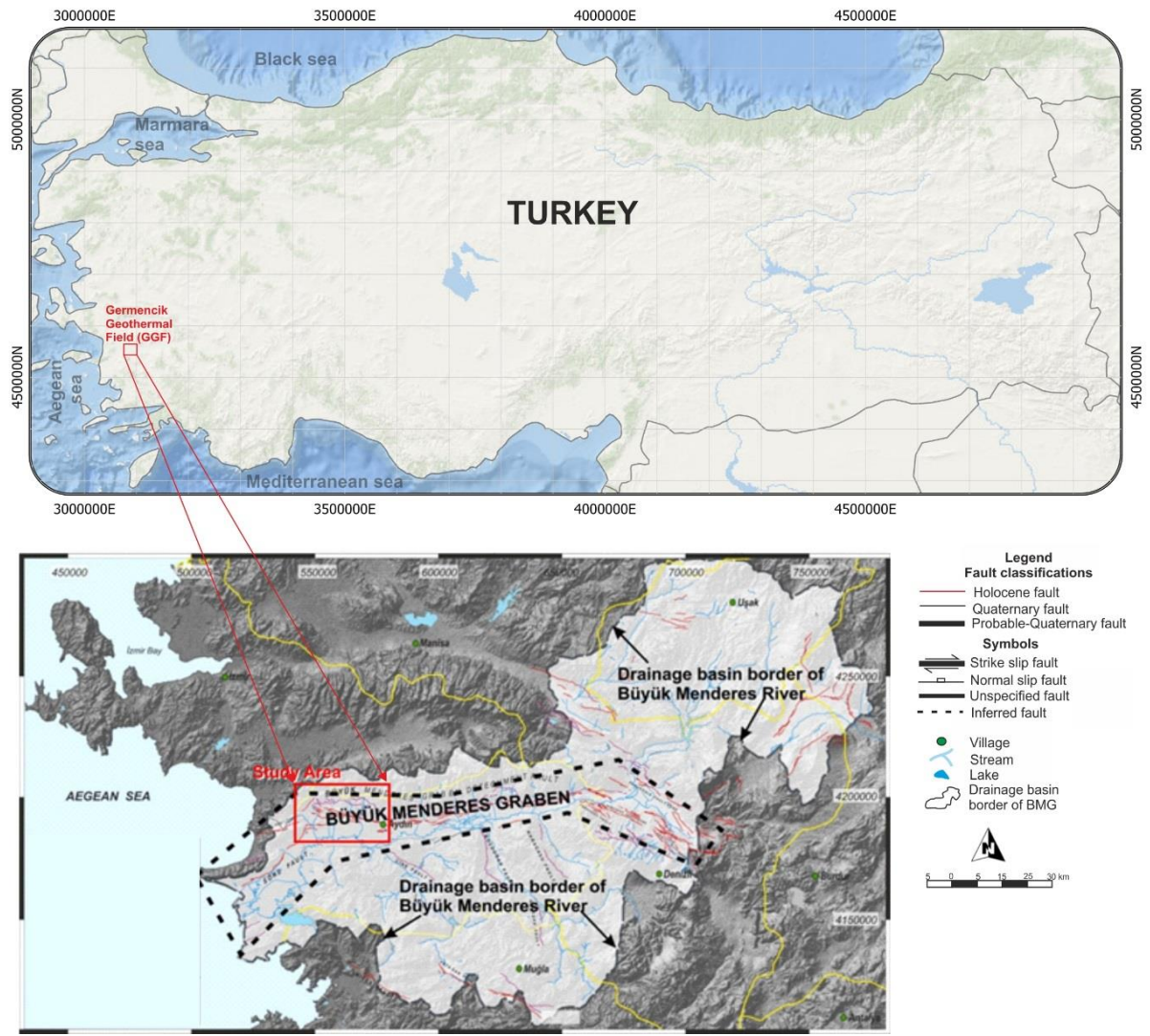

Figure 1. Location map of the study area

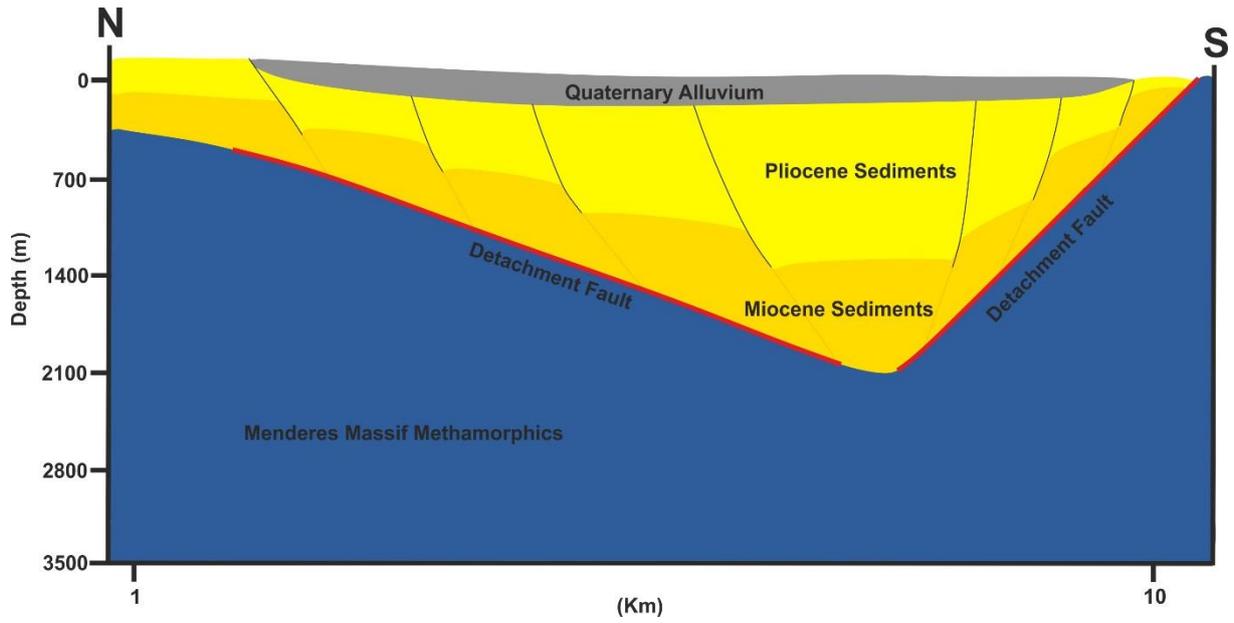

Figure 2. N-S cross section of the BMG (modified after Yamanlar et al., 2020)

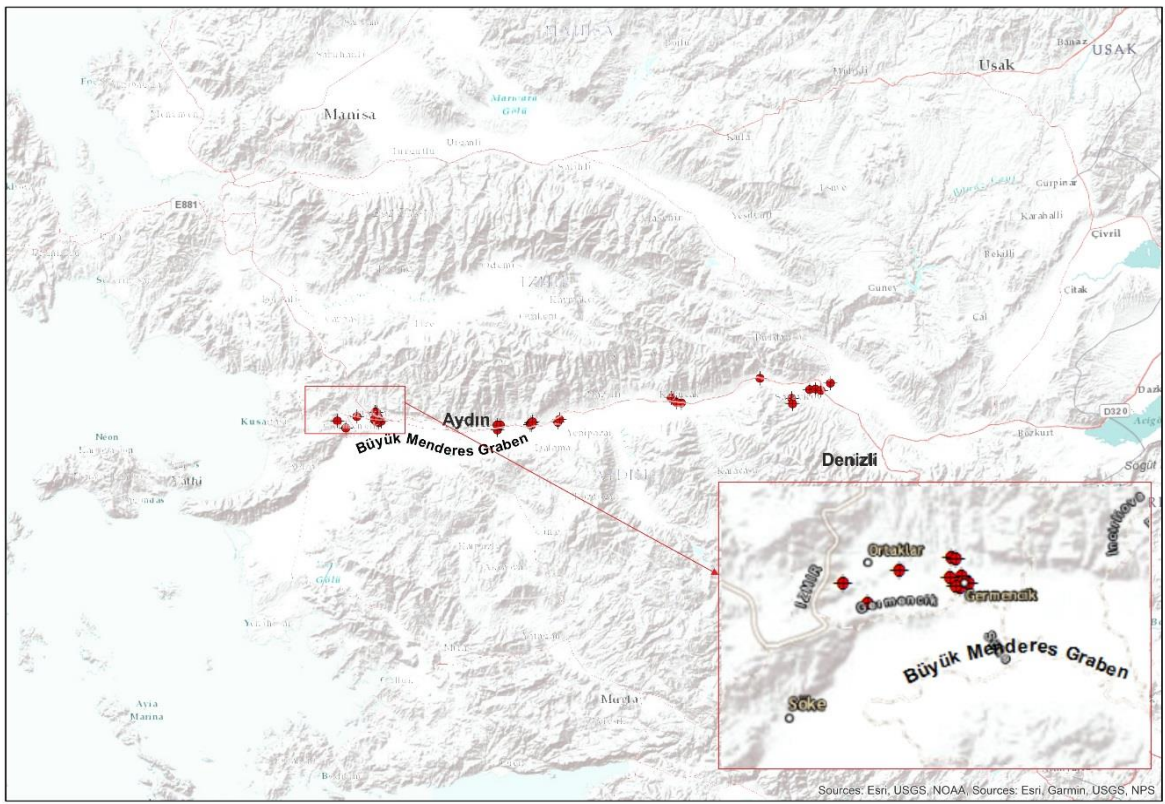

Figure 3. Geothermal power plants (red circle) in the BMG

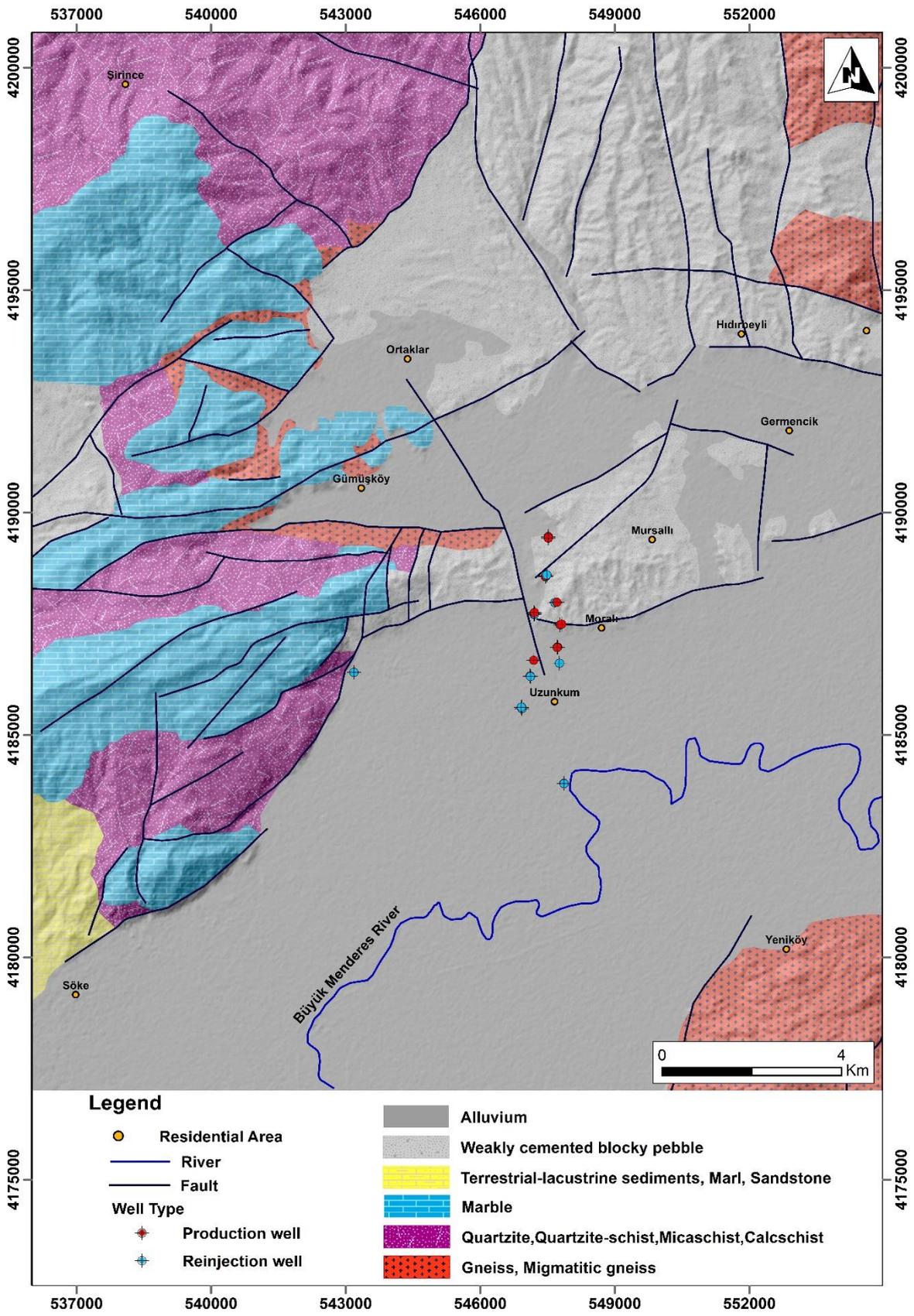

Figure 4. Geology map of the study area

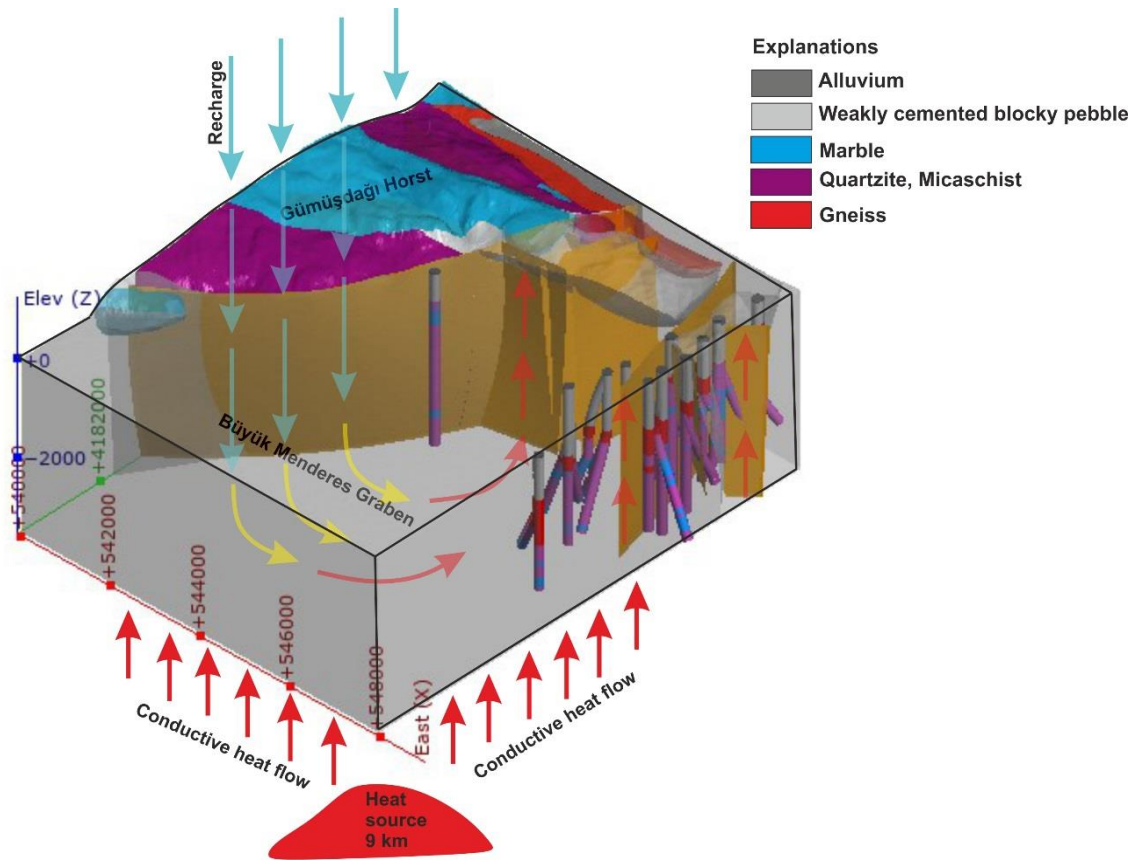
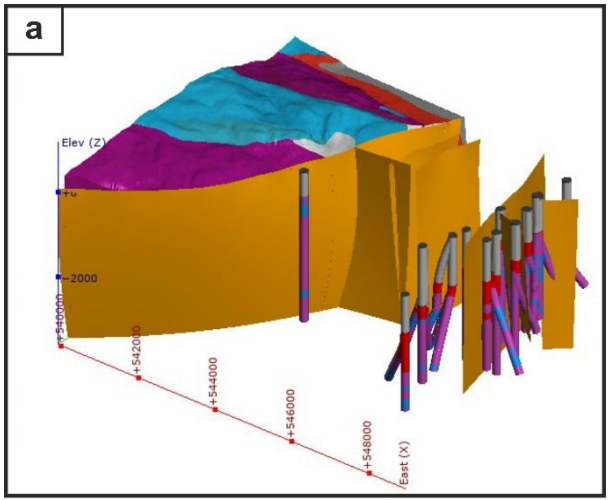
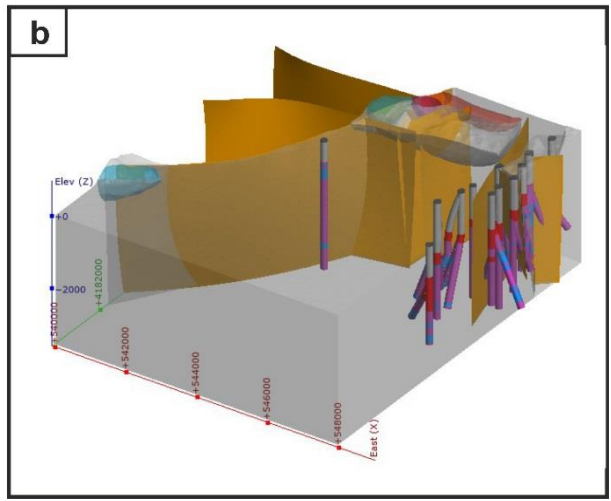

Figure 5. The 3D conceptual model of the GGF
a) West block, b) East block

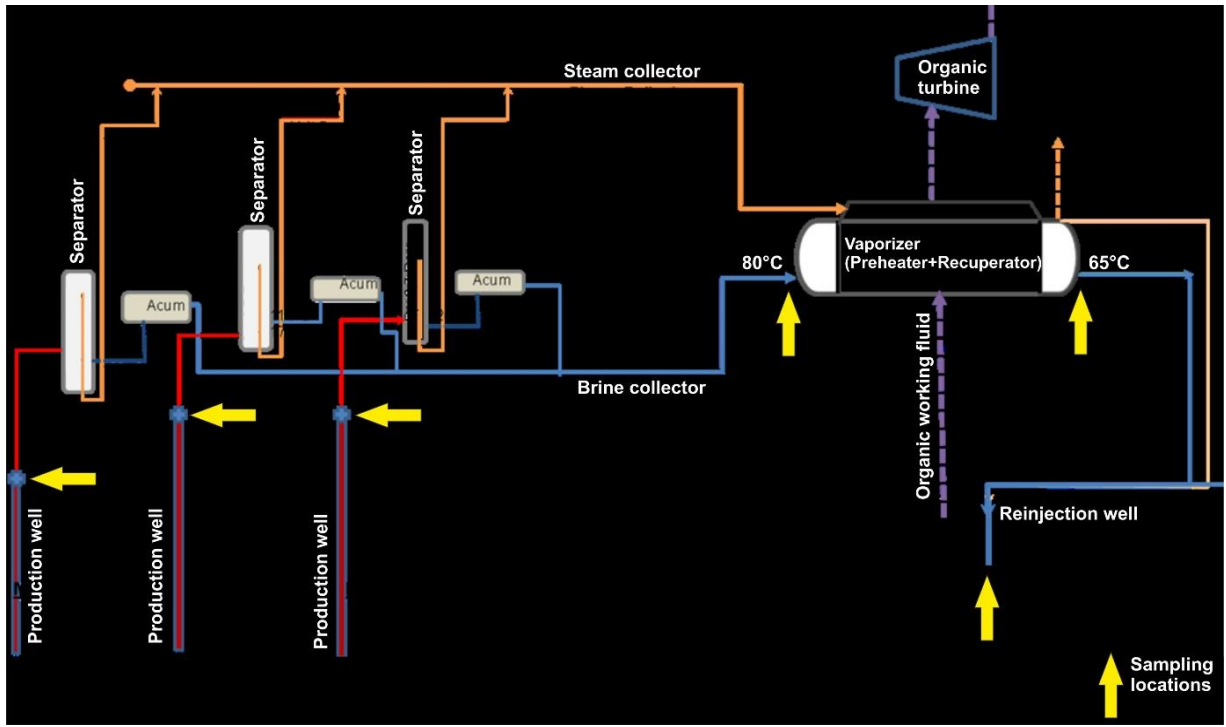
Figure 6. Schematic representation of sampling points in the GGF

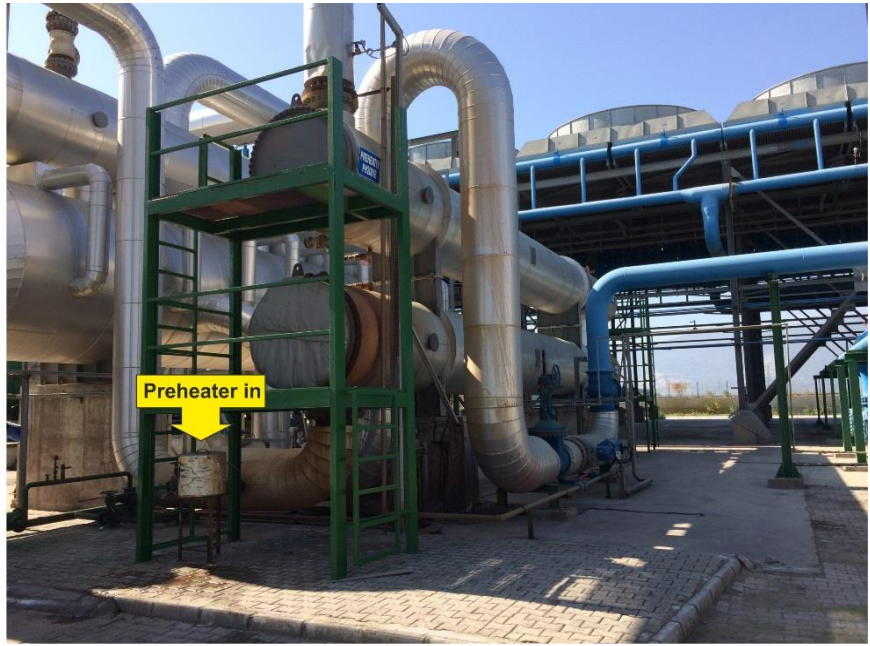

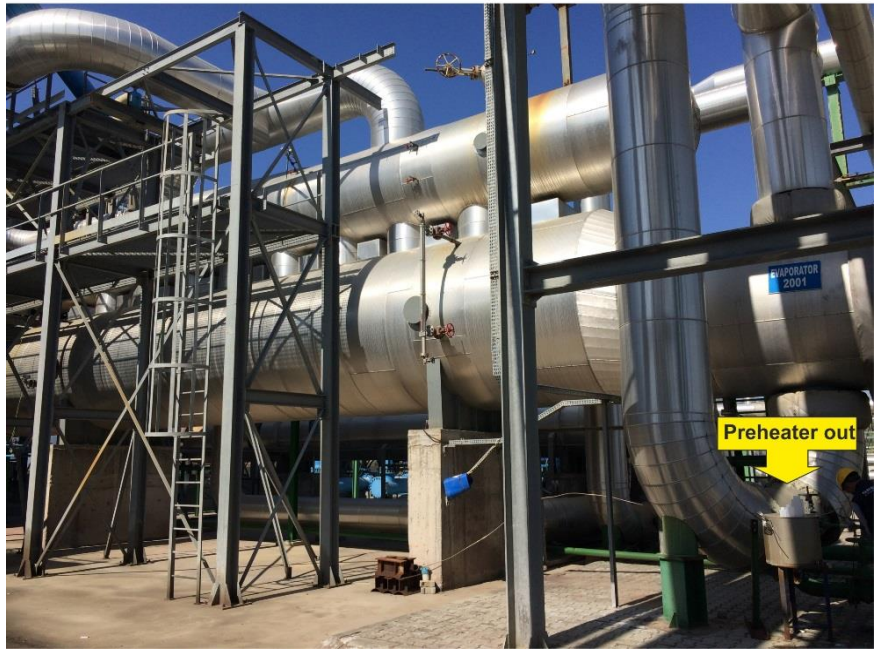

Figure 7. Sampling points in the surface equipment system of the GGF
(Photo credit: Beştepeler company)

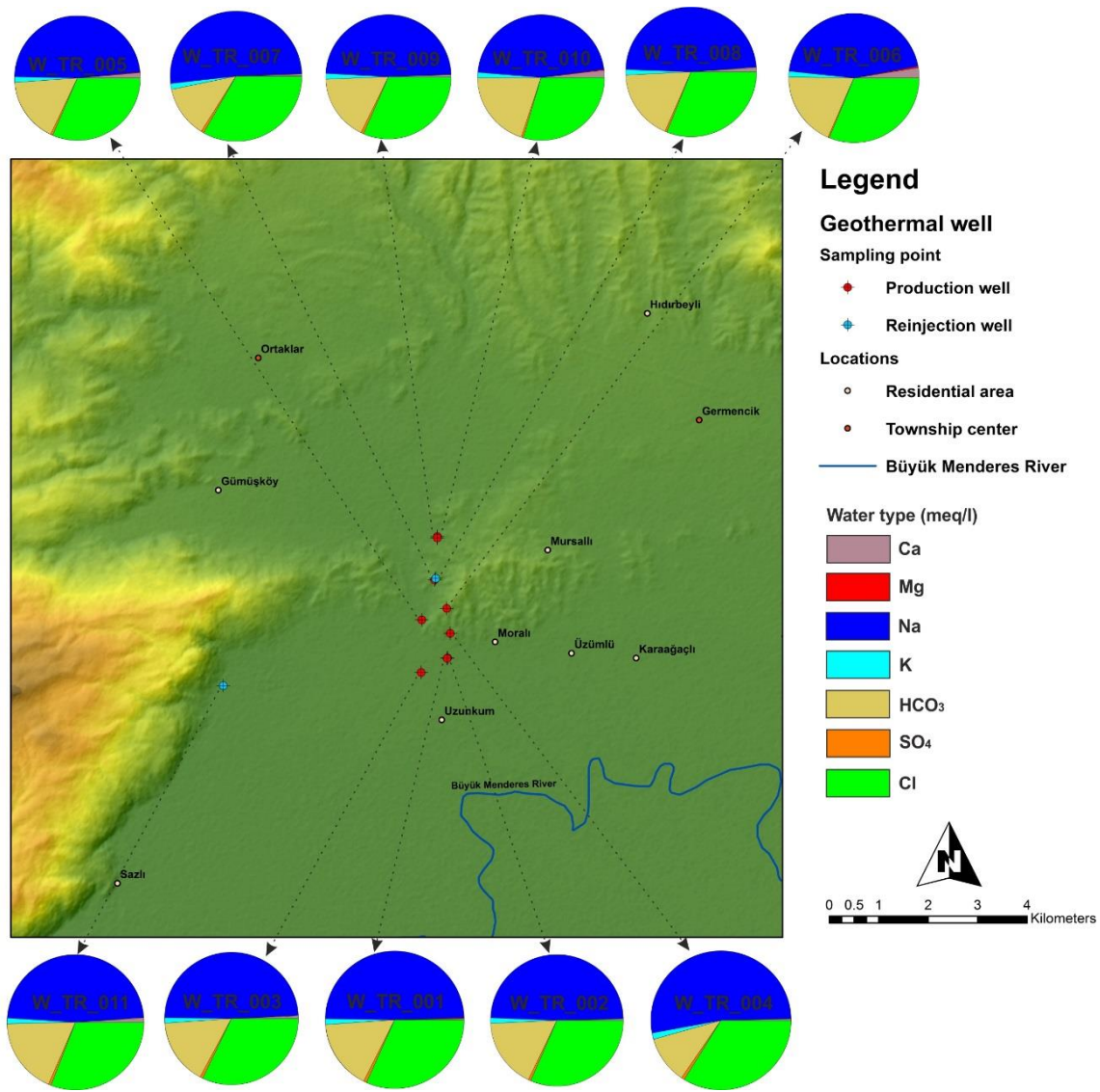

Figure 8. Demonstration of the chemical properties of the geothermal wells in the GGF

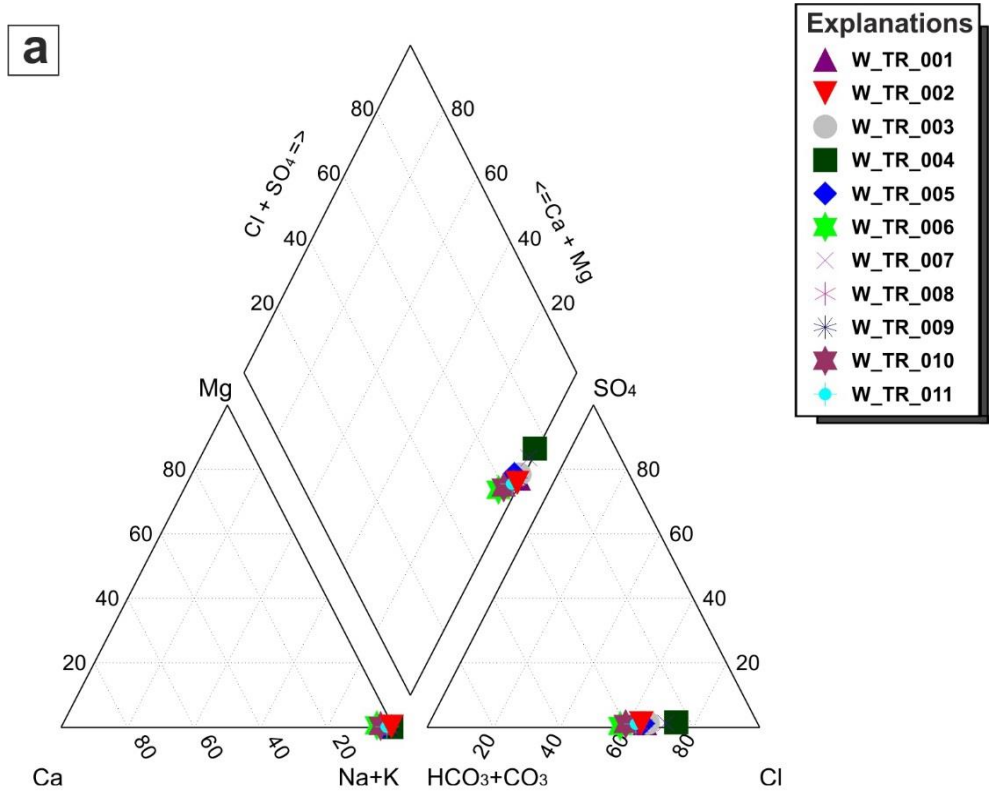

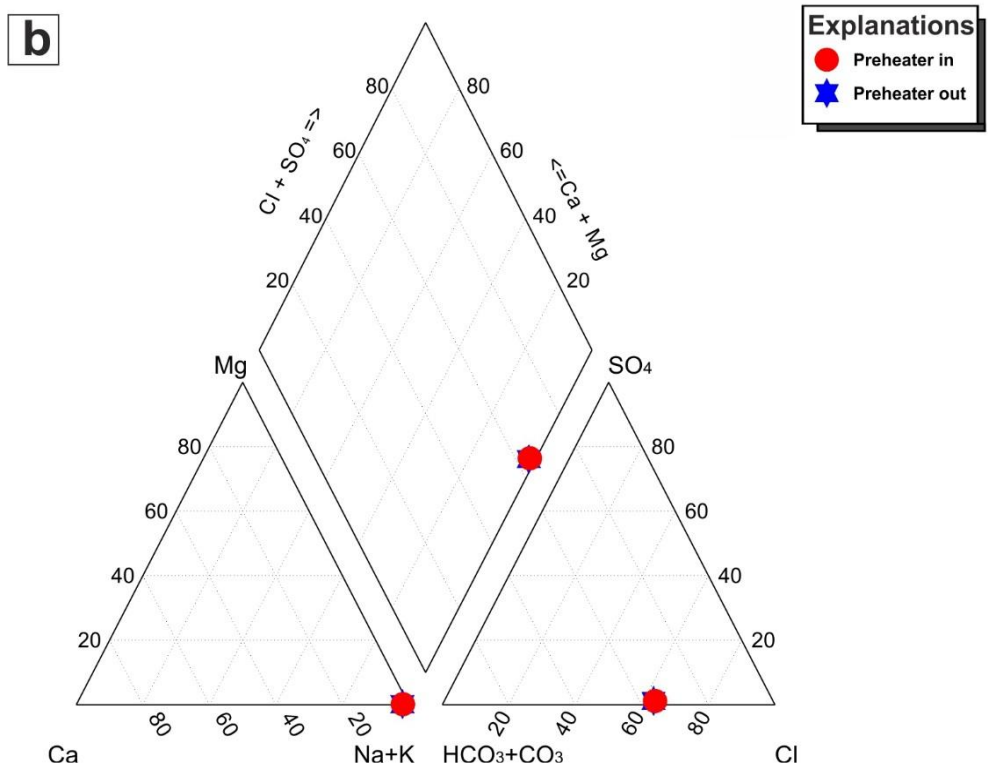

Figure 9. Piper diagrams for the geothermal waters
a) geothermal wells b) preheater system

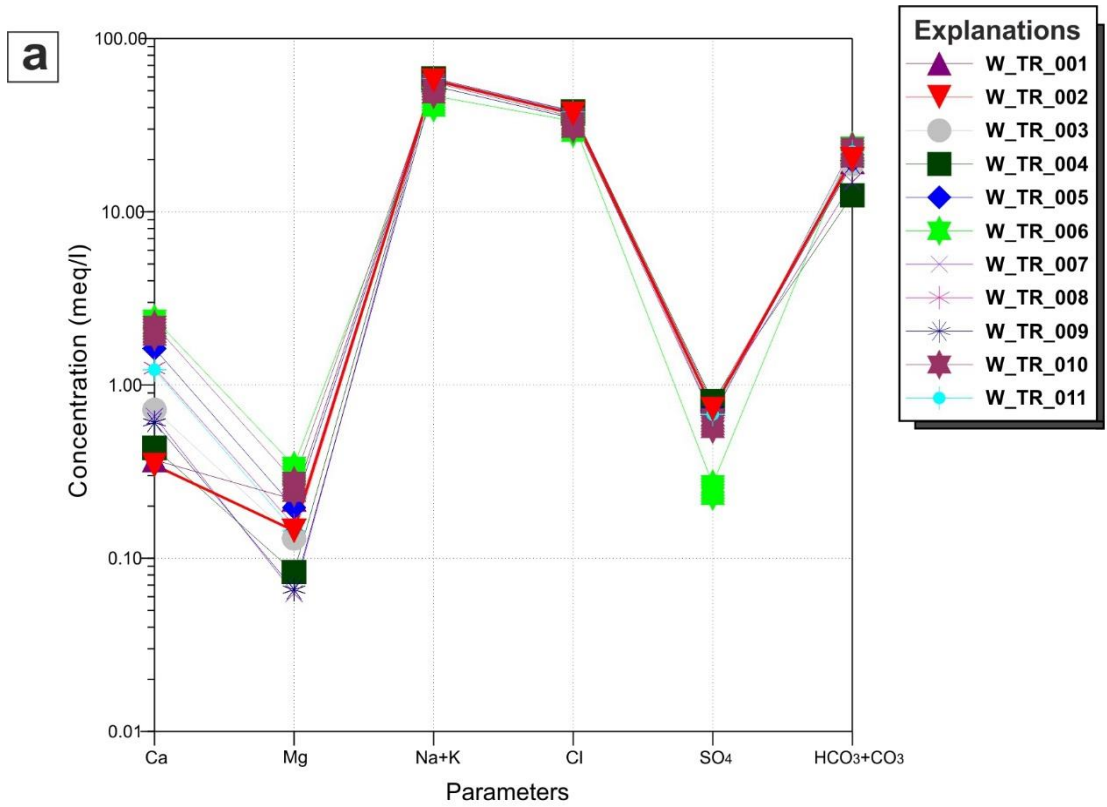

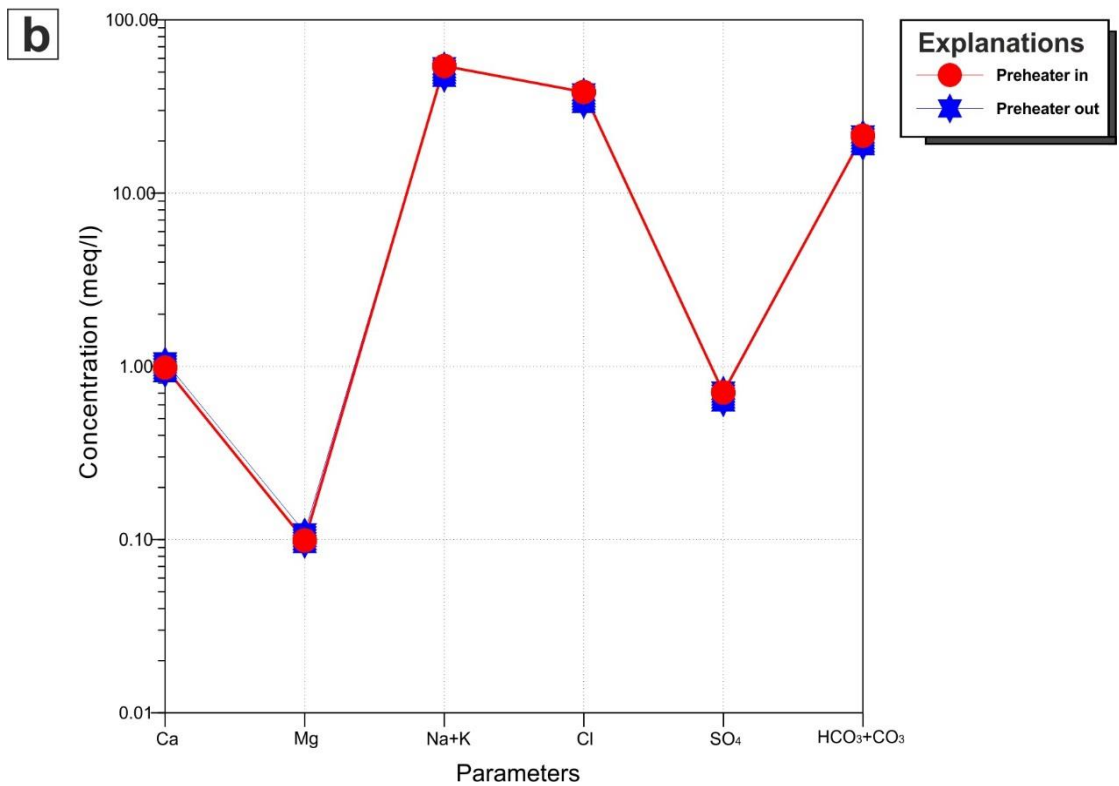

Figure 10. Schoeller diagrams for the geothermal waters
a) geothermal wells b) preheater system

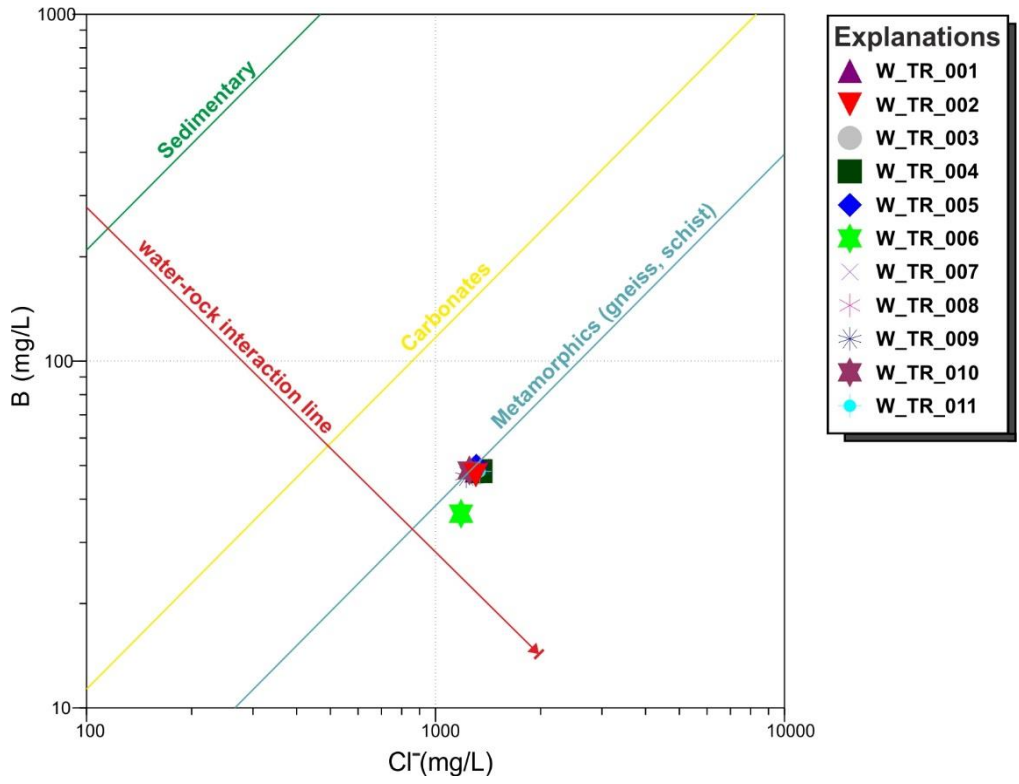

Figure 11. B and Cl⁻ relationship in the GGF fluids

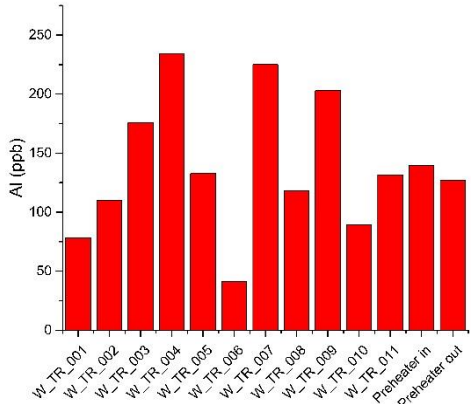
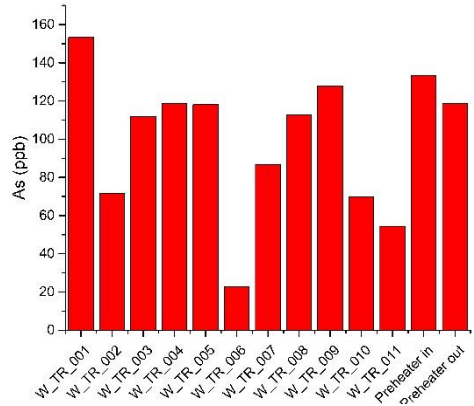
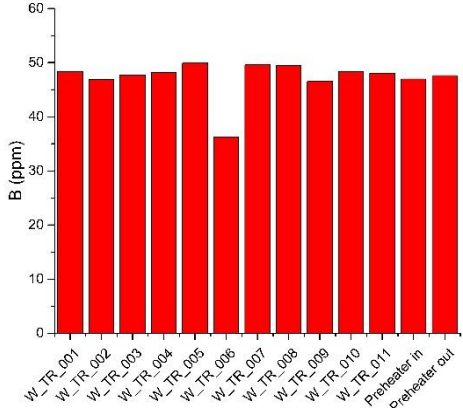
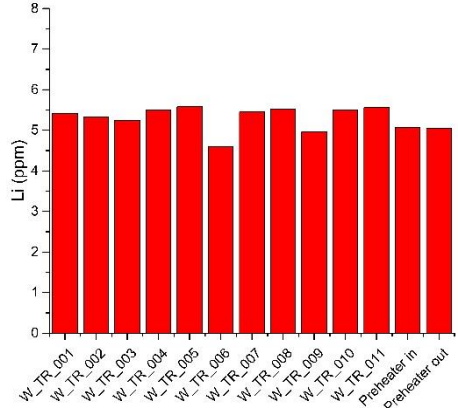
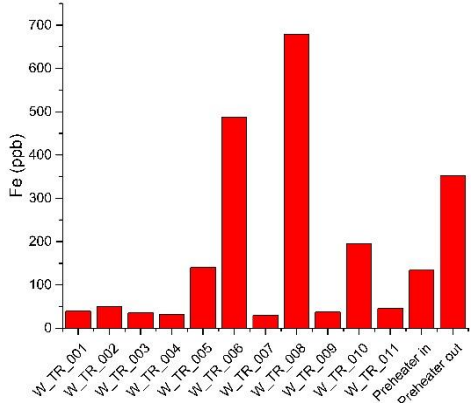
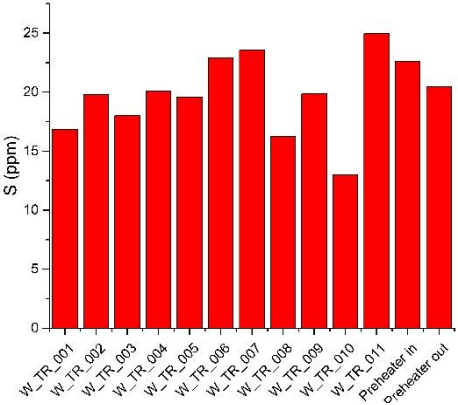

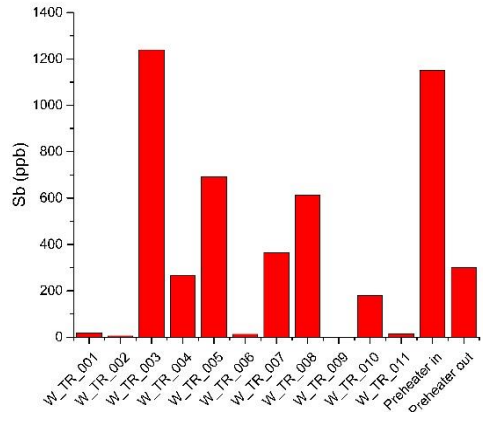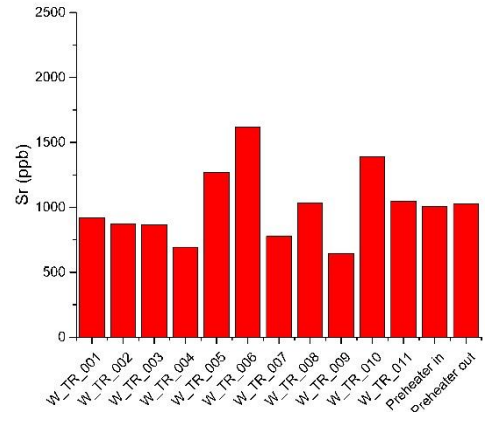

Figure 12. Heavy metal concentrations and trace elements in the geothermal waters of the GGF

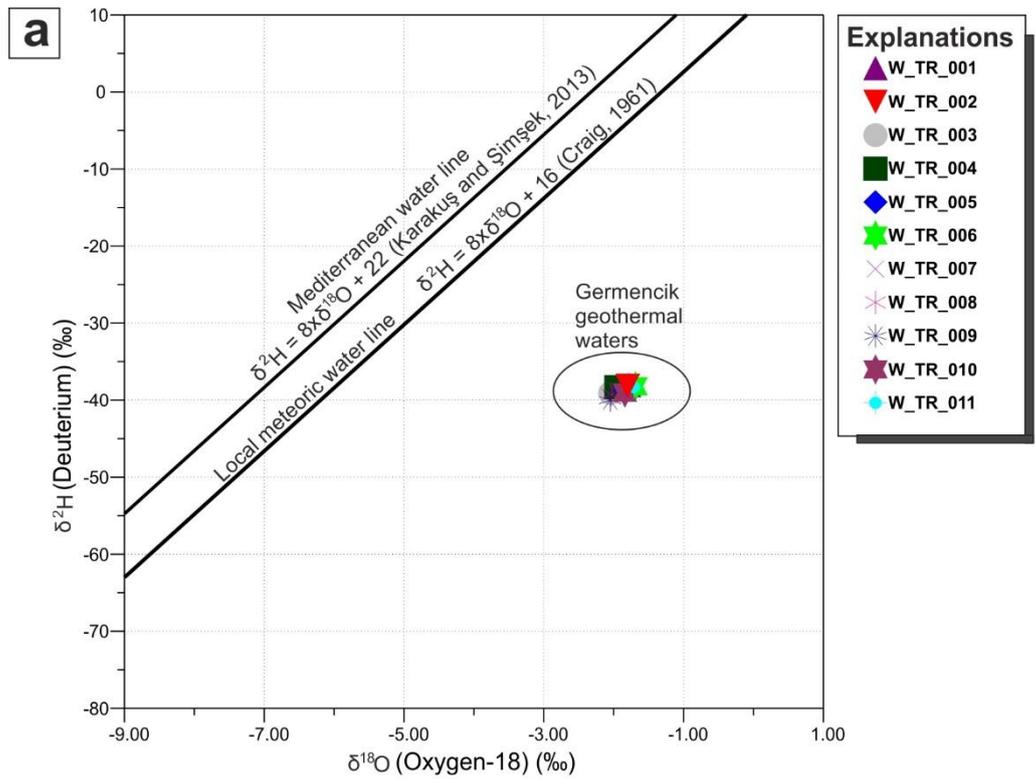
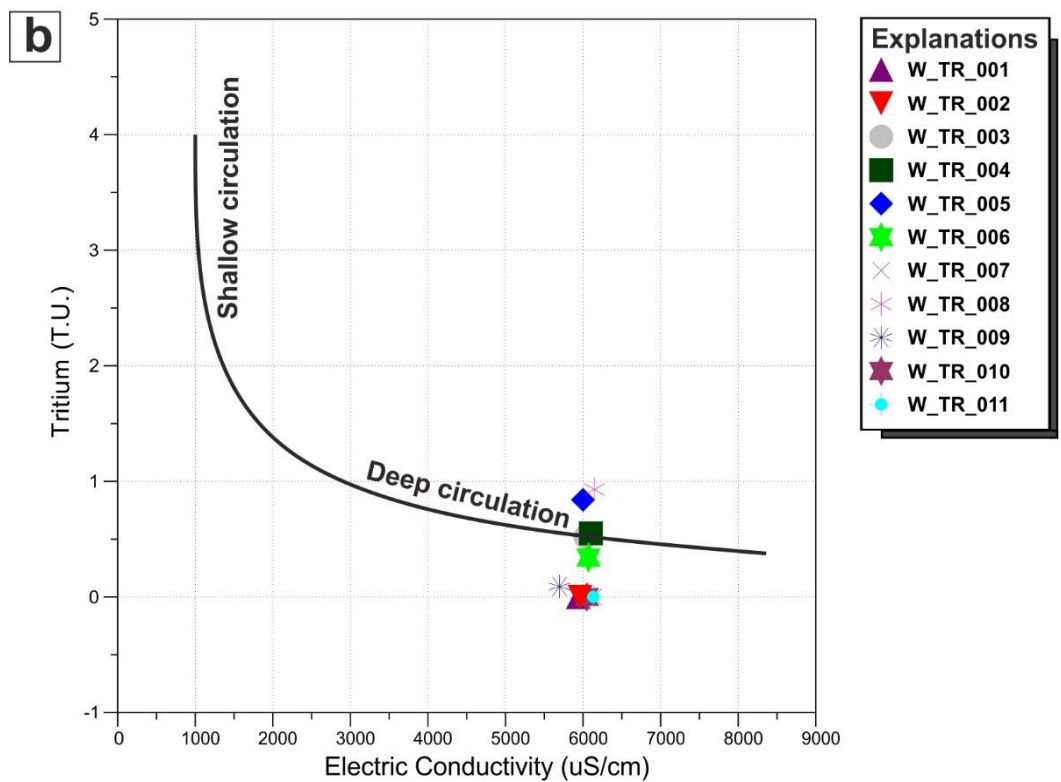

Figure 13. Isotope ratios in the groundwaters of the GGF
a) $^2$H *vs* $^{18}$O graph for the geothermal waters in the GGF
b) Tritium *vs* electric conductivity graph for the geothermal waters in the GGF

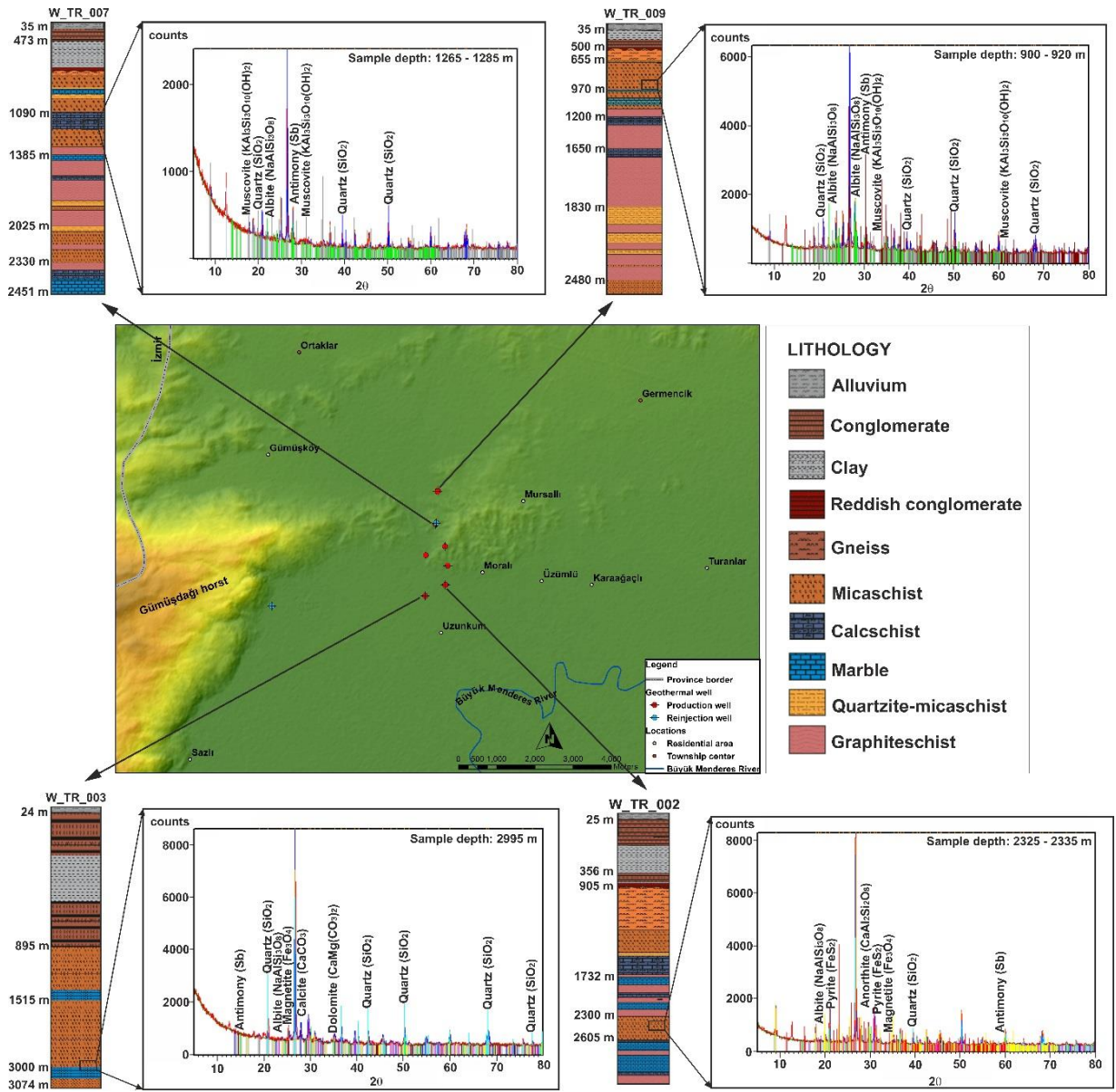

Figure 14. XRD results for the rock sample in the wells

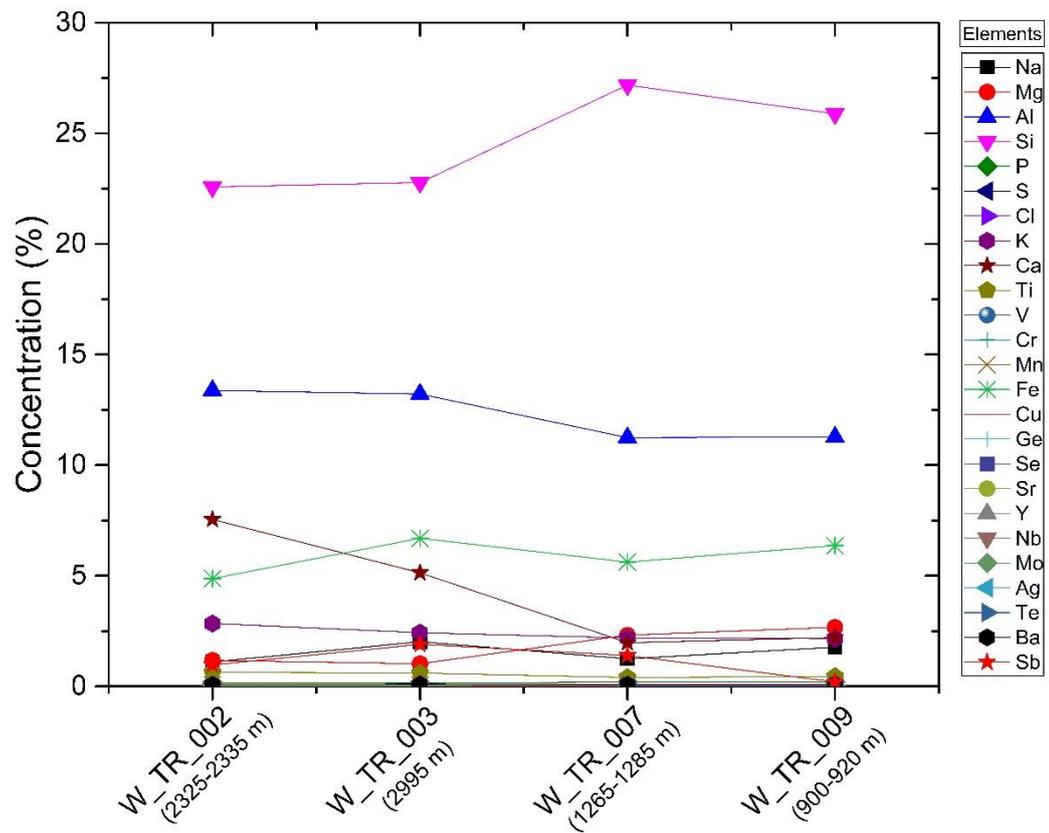

Figure 15. Elemental composition of rock samples in the well obtained by XRF as a function of depth the samples collected. The rock samples were collected from the geothermal wells.

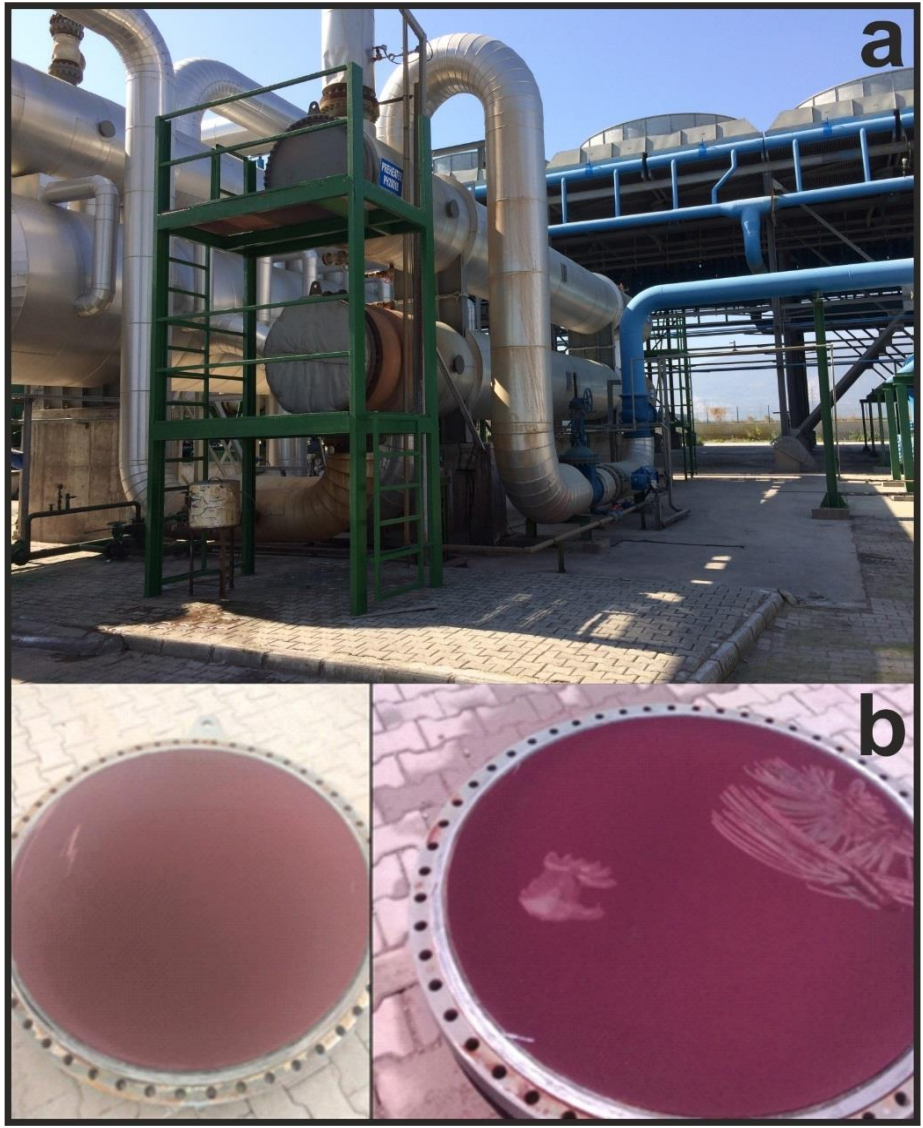

Figure 16. Sb scale encountered in the preheater in the ORC-binary system in the GGF
a)   Preheater in, b) Sb scale in the preheater system
(Photo credit: Beştepeler company)

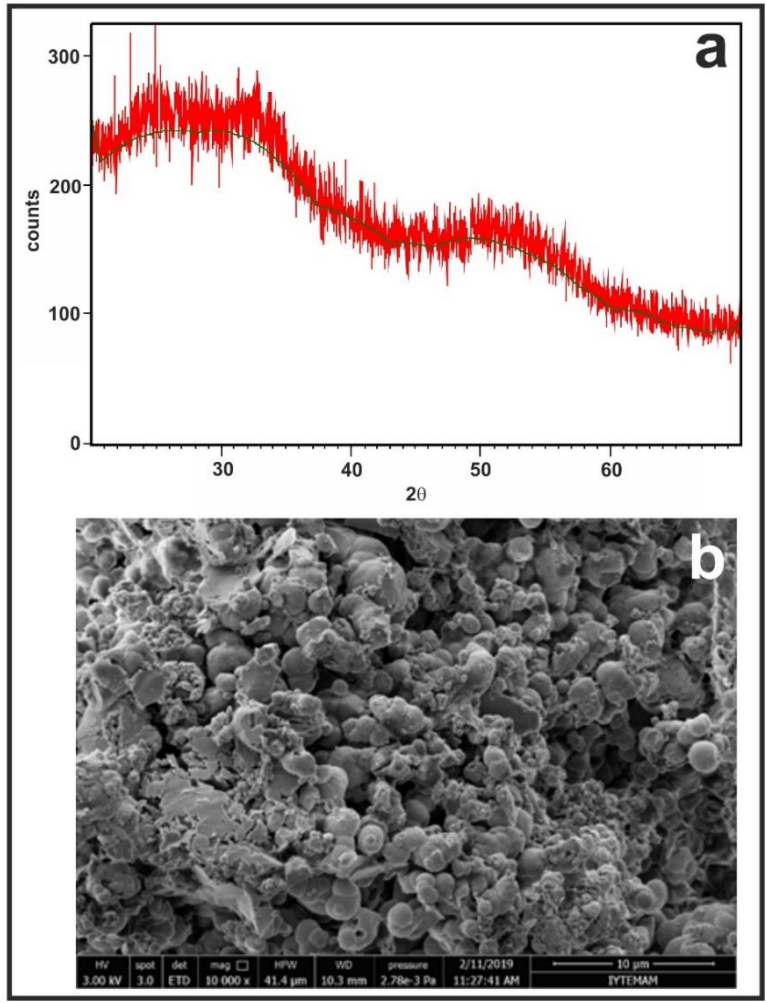

Figure 17. Evaluation of the Sb scale
a) XRD result for the scale sample, b) SEM images of the Sb deposits

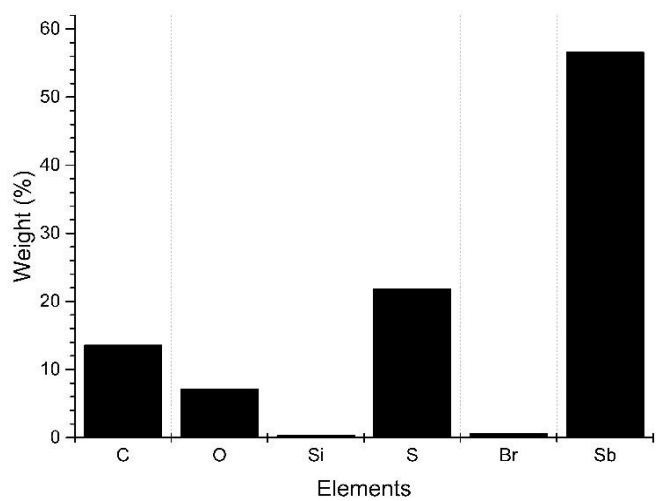

Figure 18. Elemental composition of the scale sample

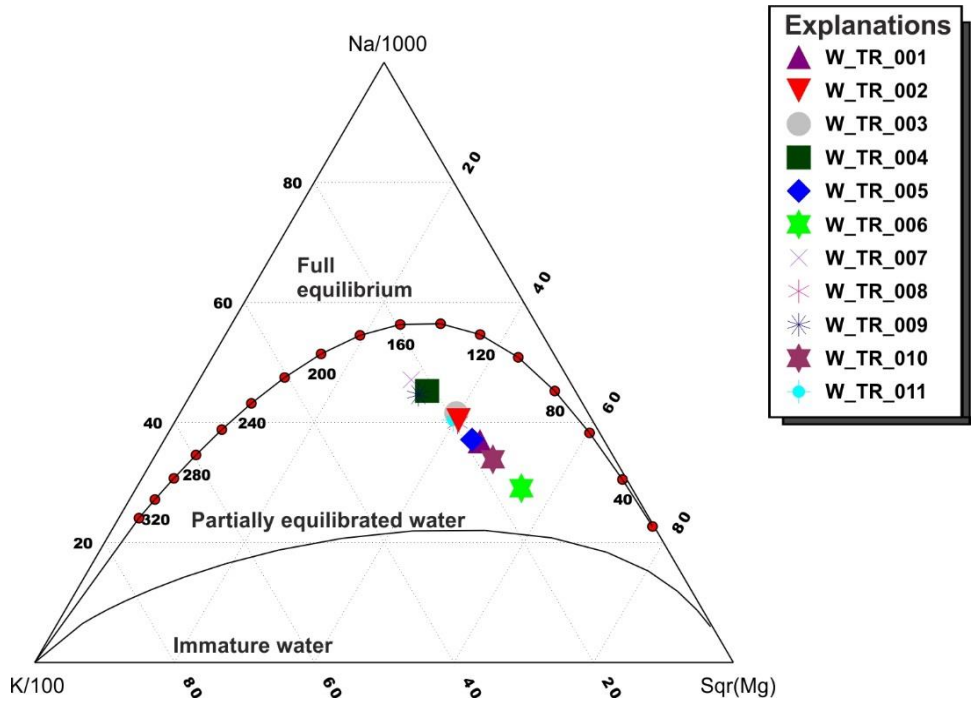

Figure 19. Distribution of water samples from the GGF in a Na–K–Mg triangular diagram

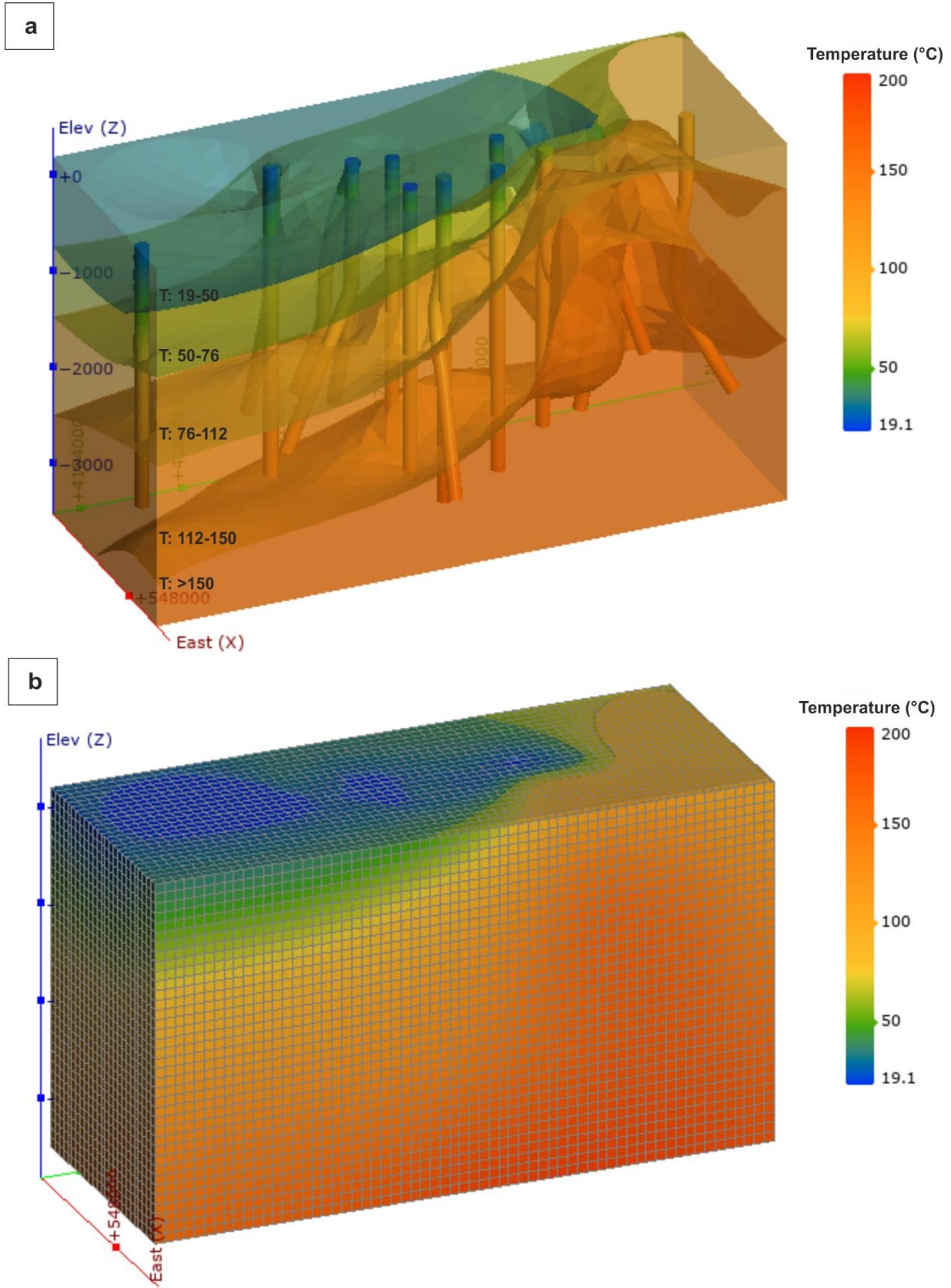

Figure 20. Reservoir temperature modelling
a) Numeric model, b) Block model

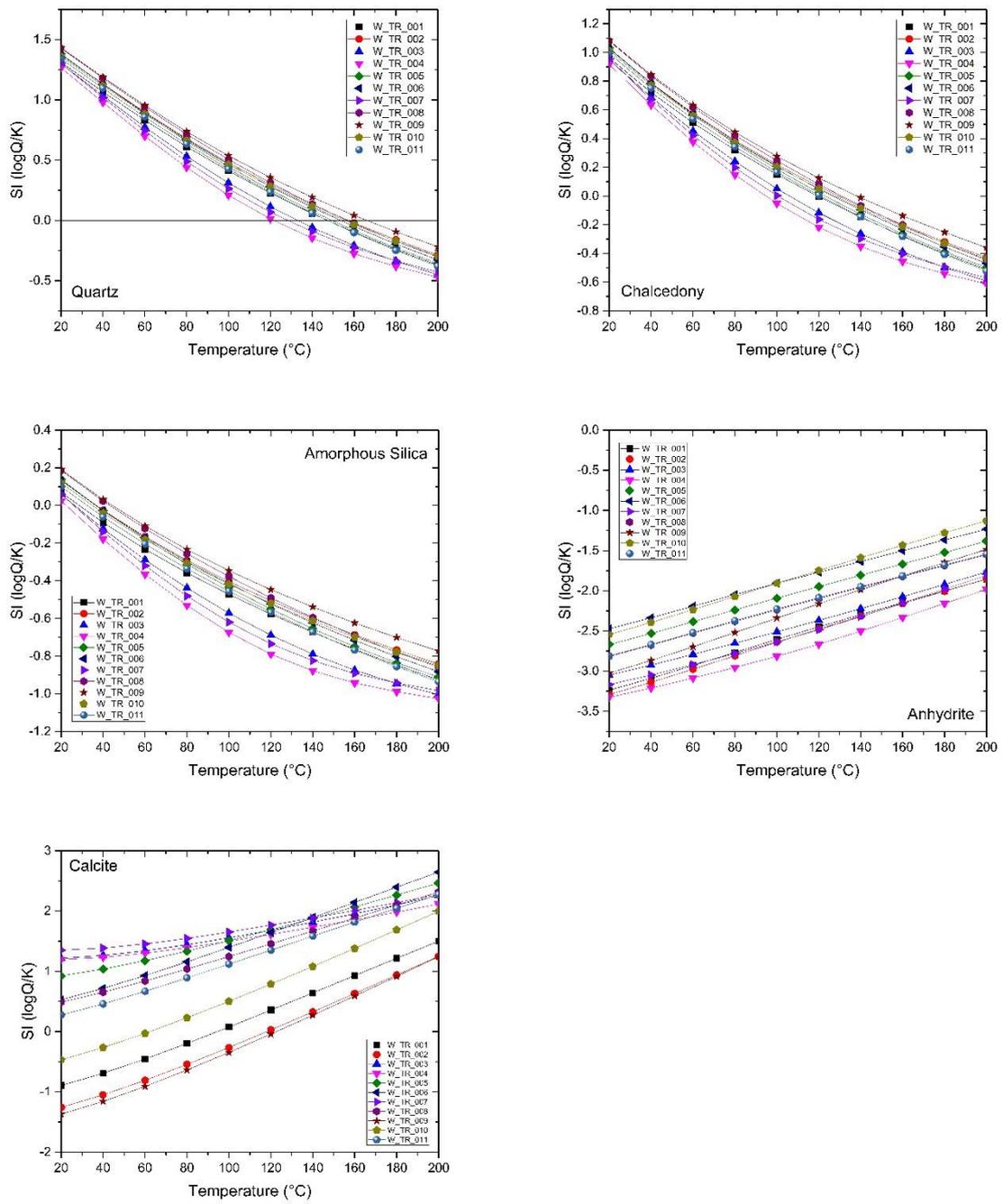

Figure 21. Mineral saturation-temperature diagrams for the downholes in the GGF

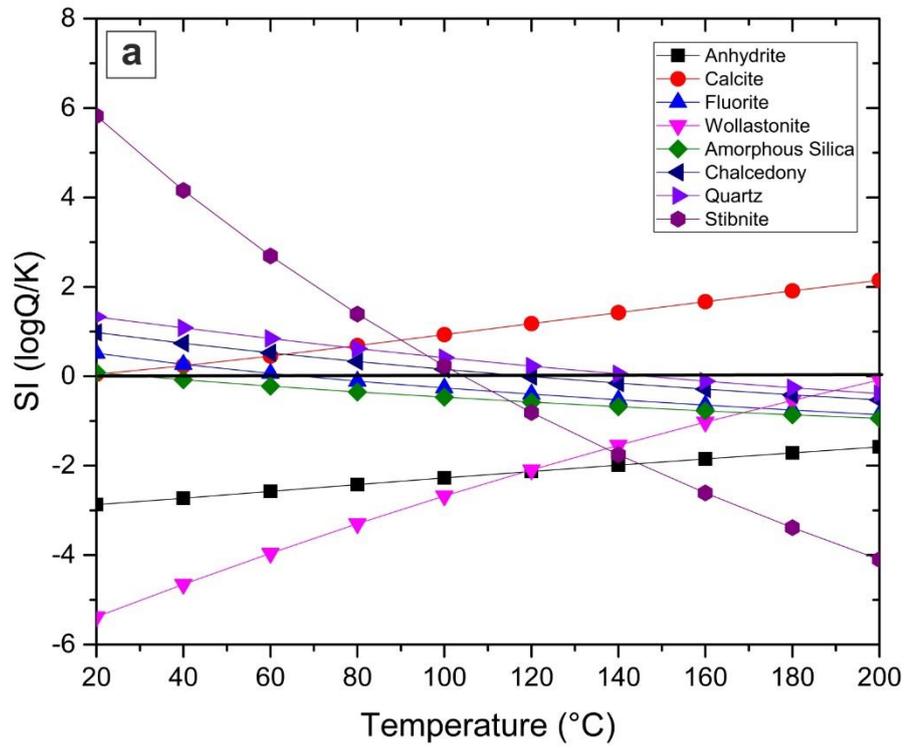

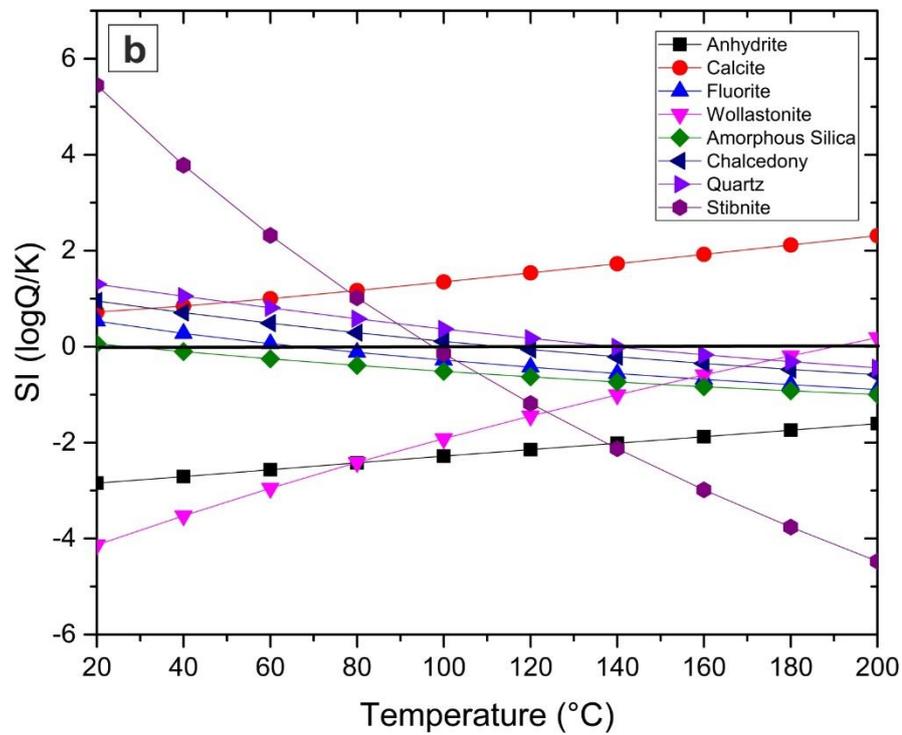

Figure 22. Mineral saturation-temperature diagrams for the preheater system in the GGF
a) Preheater in, b) Preheater out

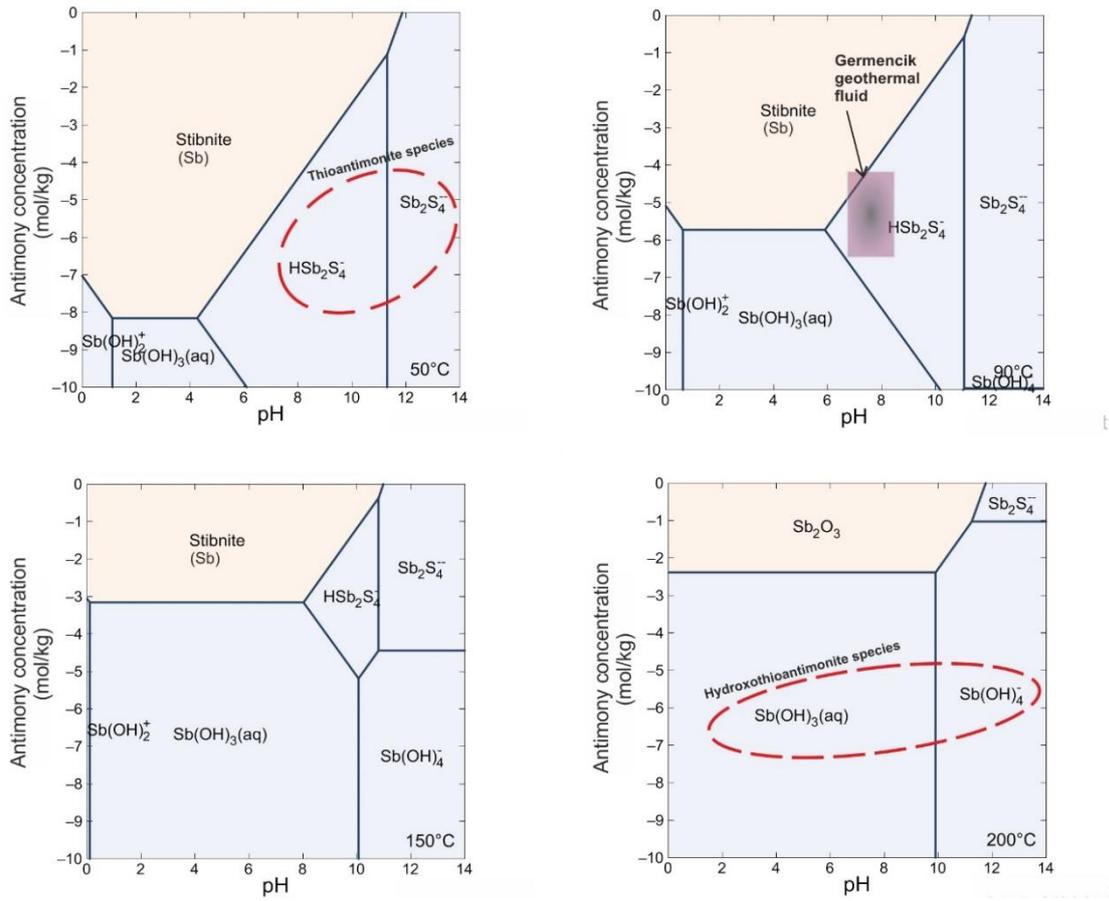

Figure 23. Stibnite species diagrams for different temperatures

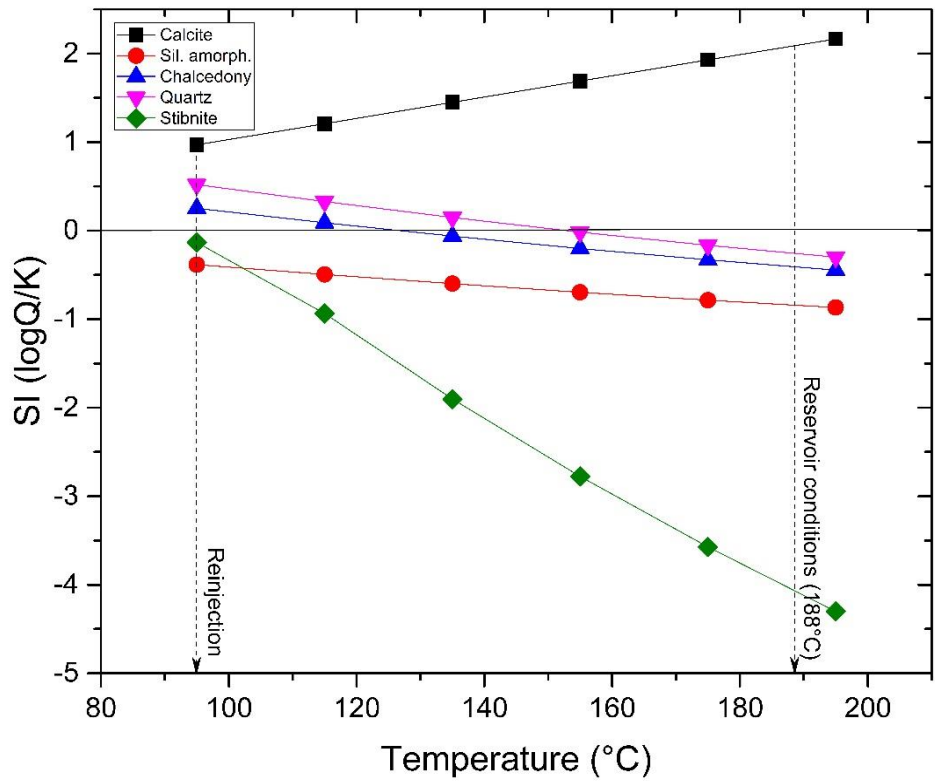

Figure 24. Critical mineral precipitation tendencies from production to reinjection in the GGF

Table 1. Rock core sampling depths and XRF measurement conditions

| Well ID | Sampling Method | Sample depth (m) | Rock type | XRF sample mass (g) | Sample state | Sample rotation |
|---|---|---|---|---|---|---|
| **W_TR_002** | Core sample | 1145-1170 | Gneiss | 0.96 | powder | no |
| | | 2140-2190 | Marble | 0.83 | powder | no |
| | | 2325-2335 | Quartzite | 1.06 | powder | no |
| | | 2670-2685 | Marble | 1.01 | powder | no |
| | | 2855-2865 | Marble | 0.84 | powder | no |
| **W_TR_003** | Core sample | 1225 | Schist | 1.10 | powder | no |
| | | 1665-1675 | Schist | 0.95 | powder | no |
| | | 2235 | Quartzite | 1.13 | powder | no |
| | | 2995 | Greenschist | 0.94 | powder | no |
| | | 3005 | Schist | 1.05 | powder | no |
| **W_TR_007** | Core sample | 1000-1040 | Quartzite | 1.35 | powder | no |
| | | 1245-1250 | Quartzite | 1.07 | powder | no |
| | | 1265-1285 | Quartzite | 1.01 | powder | no |
| | | 1960-1965 | Schist | 0.93 | powder | no |
| | | 2380-2385 | Schist | 1.00 | powder | no |
| **W_TR_009** | Core sample | 600-625 | Gneiss | 1.02 | powder | no |
| | | 900-920 | Quartzite | 1.13 | powder | no |
| | | 1000-1010 | Marble | 1.08 | powder | no |
| | | 1910-1995 | Quartzite | 1.24 | powder | no |
| | | 2100-2170 | Quartzite | 1.08 | powder | no |
| | | 2200-2250 | Schist | 1.04 | powder | no |
| | | 2450-2530 | Quartzite | 1.03 | powder | no |

Table 2. Physical properties of the geothermal water samples

| No | Well ID | pH | EC µS/cm | T (°C) | TDS (ppm) | Depth (m) | Well type |
|---|---|---|---|---|---|---|---|
| 1 | W_TR_001 | 7.1 | 5937 | 125.3 | 3700 | 3328 | Production |
| 2 | W_TR_002 | 6.87 | 5961 | 134.36 | 3900 | 3165 | Production |
| 3 | W_TR_003 | 8.24 | 6030 | 140.06 | 3700 | 3074 | Production |
| 4 | W_TR_004 | 8.54 | 6101 | 140.57 | 4000 | 3135 | Production |
| 5 | W_TR_005 | 7.82 | 5999 | 138.41 | 3900 | 3135 | Production |
| 6 | W_TR_006 | 6.25 | 6070 | 132.13 | 4200 | 1449 | Production |
| 7 | W_TR_007 | 8.47 | 6131 | 145.76 | 4100 | 2451 | Production |
| 8 | W_TR_008 | 7.62 | 6147 | 64.83 | 3900 | 2907 | Reinjection |
| 9 | W_TR_009 | 6.7 | 5697 | 148.09 | 3900 | 2568.34 | Production |
| 10 | W_TR_010 | 6.85 | 6049 | 143.65 | 3800 | 2680 | Production |
| 11 | W_TR_011 | 7.45 | 6131 | 64.51 | 4000 | 3525 | Reinjection |
| 12 | Preheater in | 7.35 | 6074 | 80 | 3900 | - | Surface equipment |
| 13 | Preheater out | 7.85 | 6090 | 65 | 4000 | - | Surface equipment |

Table 3. Chemical properties of the geothermal water samples (mg/l)

| No | Well ID | $Ca^{2+}$ | $Mg^{2+}$ | $Na^+$ | $K^+$ | $Cl^-$ | $SO_4^{2-}$ | $HCO_3^-$ | $SiO_2$ | Water Type |
|---|---|---|---|---|---|---|---|---|---|---|
| 1 | W_TR_001 | 7.40 | 2.64 | 1324.56 | 64.03 | 1312.33 | 37.41 | 1180.21 | 125 | Na-Cl-$HCO_3$ |
| 2 | W_TR_002 | 6.89 | 1.75 | 1319.22 | 62.64 | 1305.01 | 34.86 | 1230.54 | 145 | Na-Cl-$HCO_3$ |
| 3 | W_TR_003 | 14.41 | 1.58 | 1319.41 | 59.97 | 1328.42 | 38.12 | 1148.03 | 140 | Na-Cl-$HCO_3$ |
| 4 | W_TR_004 | 8.68 | 1.00 | 1355.79 | 63.59 | 1344.67 | 38.68 | 759.72 | 155 | Na-Cl-$HCO_3$ |
| 5 | W_TR_005 | 32.53 | 2.37 | 1301.49 | 65.99 | 1308.80 | 34.29 | 1206.28 | 150 | Na-Cl-$HCO_3$ |
| 6 | W_TR_006 | 48.29 | 4.15 | 1072.65 | 58.43 | 1183.42 | 12.85 | 1468.41 | 145 | Na-Cl-$HCO_3$ |
| 7 | W_TR_007 | 13.19 | 0.75 | 1352.62 | 64.77 | 1352.37 | 35.5 | 904.77 | 165 | Na-Cl-$HCO_3$ |
| 8 | W_TR_008 | 25.08 | 1.88 | 1352.98 | 65.92 | 1340.18 | 31.63 | 1313.15 | 165 | Na-Cl-$HCO_3$ |
| 9 | W_TR_009 | 12.17 | 0.79 | 1224.32 | 62.44 | 1225.77 | 34.25 | 1153.21 | 165 | Na-Cl-$HCO_3$ |
| 10 | W_TR_010 | 44.56 | 3.39 | 1281.19 | 66.22 | 1250.75 | 31.22 | 1447.48 | 140 | Na-Cl-$HCO_3$ |
| 11 | W_TR_011 | 24.53 | 1.74 | 1345.58 | 66.38 | 1337.69 | 32.47 | 1356.18 | 135 | Na-Cl-$HCO_3$ |
| 12 | Preheater in | 19.71 | 1.20 | 1244.92 | 64.01 | 1357.74 | 33.89 | 1307.45 | 130 | Na-Cl-$HCO_3$ |
| 13 | Preheater out | 21.39 | 1.33 | 1244.03 | 63.33 | 1354.89 | 34.76 | 1331.42 | 125 | Na-Cl-$HCO_3$ |

Table 4. The silica and cation geothermometer equations used in this study

| No | Geothermometer | Geothermometer Equations | Reference |
|---|---|---|---|
| 1 | Quartz | t°C=-42.2+0.28832$S$-3.6686x10$^{-4}$ $S^2$+ 3.1665x10$^{-7}$$S^3$+77.034log$S$ | Fournier and Potter (1982) |
| 2 | Quartz (at 100°C max steam loss) | t°C=(1522/(5.75-logS))-273.15 | Fournier (1977) |
| 3 | Quartz | t°C =-55.3+0.36559$S$-5.3954x10$^{-4}$ $S^2$ + 5.5132x10$^{-7}$$S^3$+74.360log$S$ | Arnórsson (2000) |
| 4 | Na-K | t°C =(1319/(1.699+log(Na/K)))-273.15 | Arnorsson et al. (1983) |
| 5 | Na-K | t°C =(1217/(1.483+log(Na/K)))-273.15 | Fournier and Potter (1979) |
| 6 | Na-K | t°C =(1178/(1.470+log(Na/K)))-273.15 | Nieva and Nieva (1987) |
| 7 | Na-K | t°C =(1390/(1.750+log(Na/K)))-273.15 | Giggenbach (1988) |

Table 5. Reservoir temperatures (°C) calculated with silica and cation geothermometers of the GGF

| Location | Well ID | Geothermometer Equations | | | | | | |
|---|---|---|---|---|---|---|---|---|
| | | 1 | 2 | 3 | 4 | 5 | 6 | 7 |
| Germencik Geothermal Field (GGF) | W_TR_001 | 150 | 143 | 139 | 164 | 162 | 150 | 180 |
| | W_TR_002 | 159 | 151 | 149 | 163 | 160 | 149 | 179 |
| | W_TR_003 | 157 | 149 | 146 | 161 | 158 | 146 | 176 |
| | W_TR_004 | 163 | 154 | 153 | 162 | 160 | 148 | 178 |
| | W_TR_005 | 161 | 153 | 151 | 167 | 165 | 153 | 183 |
| | W_TR_006 | 159 | 151 | 149 | 172 | 170 | 158 | 188 |
| | W_TR_007 | 167 | 158 | 158 | 164 | 161 | 149 | 180 |
| | W_TR_008 | 167 | 158 | 158 | 165 | 162 | 150 | 181 |
| | W_TR_009 | 167 | 158 | 158 | 168 | 165 | 153 | 184 |
| | W_TR_010 | 157 | 149 | 146 | 169 | 166 | 154 | 185 |
| | W_TR_011 | 154 | 147 | 144 | 166 | 163 | 151 | 182 |